\documentclass[12pt]{article}
\usepackage{amsmath,amsfonts,amssymb}

\begin{document}
\newcommand{\todo}[1]{{\em \small {#1}}\marginpar{$\Longleftarrow$}}
\newcommand{\labell}[1]{\label{#1}\qquad_{#1}} %{\label{#1}} %
\numberwithin{equation}{section}

\rightline{DCPT-09/45}
\vskip 1cm

\begin{center} {\Large \bf Holographic stress tensor for
    non-relativistic theories}
\end{center}
\vskip 1cm

\renewcommand{\thefootnote}{\fnsymbol{footnote}}
\centerline{\bf Simon F. Ross$^a$\footnote{S.F.Ross@durham.ac.uk} and Omid Saremi$^b$\footnote{omid@hep.physics.mcgill.ca}}
\vskip .5cm
\centerline{\it ${}^a$ Centre for Particle Theory, Department of Mathematical
  Sciences}
\centerline{\it Durham University, South Road, Durham DH1 3LE, U.K.}
\centerline{\it ${}^b$ Department of Physics, McGill University, 3600
  University St,}
\centerline{\it Montr\'eal, Que H3A 2T8, Canada}

\setcounter{footnote}{0}
\renewcommand{\thefootnote}{\arabic{footnote}}

\begin{abstract}
  We discuss the calculation of the field theory stress tensor from
  the dual geometry for two recent proposals for gravity duals of
  non-relativistic conformal field theories. The first of these has a
  Schr\"odinger symmetry including Galilean boosts, while the second
  has just an anisotropic scale invariance (the Lifshitz case). For
  the Lifshitz case, we construct an appropriate action principle. We
  propose a definition of the non-relativistic stress tensor complex
  for the field theory as an appropriate variation of the action in
  both cases. In the Schr\"odinger case, we show that this gives
  physically reasonable results for a simple black hole solution and
  agrees with an earlier proposal to determine the stress tensor from
  the familiar AdS prescription. In the Lifshitz case, we solve the
  linearised equations of motion for a general perturbation around the
  background, showing that our stress tensor is finite on-shell.
\end{abstract}

\section{Introduction}

The use of gravitational duals to study strongly-coupled field
theories \cite{mald,magoo} has produced substantial progress in our
understanding of both vacuum correlation functions and
finite-temperature behaviour at strong coupling. The domain in which
this holographic toolbox has been put into use is remarkably
large. For instance, the hydrodynamic limit of the duality has proved
insightful in studying the quark-gluon plasma created at RHIC \cite{skenderis,
  gubser, Rangamanirev}. There have also been attempts to model
interesting condensed matter systems using a corresponding
gravitational dual \cite{Hartnoll,Herzog,Kovtun,Hartnoll:2009sz}. Much
of this work has concerned relativistic theories with a conformal
symmetry in the ultraviolet, which are described by asymptotically
Anti-de Sitter (AdS) spacetimes. Largely inspired by condensed matter
systems, however, this has recently been extended to consider
non-relativistic theories with an anisotropic scaling symmetry
\begin{equation} \label{e1}
t \to \lambda^z t, \quad x^i \to \lambda x^i.
\end{equation}
The case $z=2$, which is the symmetry of a free non-relativistic
theory (as the Hamiltonian is quadratic in the momenta), is often of
particular interest. Strongly-coupled theories with this scaling
symmetry arise as critical points in condensed matter systems: two
cases of interest are where the theory has a Galilean boost symmetry
together with the anisotropic scaling symmetry, forming the
Schr\"odinger symmetry group for $z=2$
\cite{Hagen:1972pd,Mehen:1999nd,Nishida:2007pj}, and where the theory
has no boost symmetry, which we will refer to as the Lifshitz case
\cite{Hornreich:1975zz,grinstein,henley}. Dual geometries realising
these symmetries as isometries were obtained for the Schr\"odinger
symmetry in \cite{son2008,Balasubramanian:2008dm} and for the Lifshitz case in
\cite{klm}.

These geometries are, of course, not asymptotically AdS. This provides
an additional motivation for studying these systems, as the
generalisation of holographic techniques to this new context may offer
new insights into the nature of the relation between quantum gravity
in asymptotically non-AdS spacetimes and the dual field theory. It
also requires the development of a new dictionary relating bulk to the
boundary quantities.

In the asymptotically AdS case, there is a well-developed framework
for calculating field theory quantities from the bulk spacetime.  The
key element in this framework is an appropriate action principle for
the bulk theory, which is finite on-shell and stationary under
variations which satisfy some asymptotic fall-off conditions. This is
constructed by adding covariant local boundary counter-terms to the
bulk action
\cite{Henningson:1998gx,Balasubramanian:1999re}. Correlation functions
for the field theory can then be obtained by considering appropriate
variations of this action with respect to the boundary data. One
important example is the expectation value of the stress tensor, which
plays a central role in the application to finite-temperature field
theory in particular. The stress tensor is obtained by variation of
the action with respect to the boundary metric
\cite{Henningson:1998gx,Balasubramanian:1999re}. This prescription has
been extensively used in the context of AdS/CFT and elsewhere. More
recently, in one particularly interesting application, it was applied
to obtain a very beautiful and direct relationship between the
dynamics of the stress tensor in the hydrodynamic regime in the field
theory and the equations of motion of the bulk gravitational theory
\cite{Bhattacharyya:2008jc,Bhattacharyya:2008xc}.

Some progress has been made in extending these aspects of the
holographic dictionary to the asymptotically Schr\"odinger case. A
black hole solution corresponding to the finite temperature grand
canonical ensemble in the field theory was constructed in
\cite{mmt,abm,heat}. An action principle for asymptotically
Schr\"odinger spaces was constructed in \cite{heat}, by adding
local covariant boundary counter-terms to the bulk action as in the AdS
case. However, a stress tensor was not successfully constructed from
the variation of this action. The asymptotically Schr\"odinger
solutions are obtained by applying a solution-generating
transformation to asymptotically AdS solutions, and it was proposed in
\cite{mmt} that the stress tensor obtained from the asymptotically AdS
solution could be re-interpreted in terms of the non-relativistic
solution. This approach was used to study the hydrodynamic regime in
this theory in \cite{hydro} by re-using the results of
\cite{Bhattacharyya:2008jc}. For the Lifshitz case, black hole
solutions were obtained in \cite{dt,mann}, and the energy of these solutions
was studied in \cite{peet}, but an action principle and stress tensor have
not yet been obtained for this theory.

To find a detailed map between bulk fluctuations and field theory
objects, one might need to find embeddings of these spacetimes in a
complete theory of quantum gravity like string theory. This was
accomplished for the Schr\"odinger case in \cite{mmt,abm,heat}. We
will not attempt to do this for the Lifshitz case here. Rather we
study simply the generalisation of the holographic dictionary at the
level of the classical gravity in the bulk.

The aim of this paper is to further develop the holographic dictionary
for asymptotically Schr\"odinger and asymptotically Lifshitz
spacetimes, focusing on the construction of one-point functions.  We
only consider the case where the boundary metric is flat; the
extension to consider more general boundary metrics, and in particular
the case where the boundary metric is a sphere, is an interesting
problem for the future. We will construct an appropriate action
principle for the Lifshitz case in section \ref{actionsec}. We then
propose a definition of the non-relativistic stress tensor complex for
the field theory which can be applied to both Lifshitz and
Schr\"odinger cases. A key element of our definition is treating the
variation of the matter fields appropriately. Our approach is strongly
inspired by \cite{him}, which showed that in the relativistic case in
the presence of arbitrary bulk matter fields, the stress tensor is
defined by considering the variation of the boundary metric holding
fixed the tangent space components of the matter fields.  We propose
to apply the same prescription to the non-relativistic
cases. Considering the variation of the boundary geometry with the
tangent space components of the matter fields fixed turns out to be
crucial to obtain a finite stress tensor. We discuss the application
of this prescription to calculate the stress tensor in the Lifshitz
case in general in section \ref{stress}, and apply these ideas to
asymptotically Schr\"odinger spacetimes in section \ref{schr}. In the
Schr\"odinger case, we show that the results obtained from our
proposal agree with those obtained from the stress tensor of the
asymptotically AdS solution following the prescription of \cite{mmt}.

In section \ref{pert}, we solve the bulk equations of motion for a
general linearised perturbation of the Lifshitz spacetime, and
calculate our stress tensor for this linearised perturbation. We find
that the stress tensor for the linearised perturbations is finite. The
finiteness of the stress tensor is an important test of our
prescription.  We solve the bulk equations of motion for the
perturbation in a series expansion in derivatives of the perturbation
along the boundary directions. In the linearised analysis, only a
finite number of orders in this expansion make finite contributions to
the boundary stress tensor.\footnote{This is different from the
  hydrodynamic analysis, where all orders in derivatives contribute,
  because we are linearising around the zero-temperature Lifshitz
  geometry of \cite{klm}, not around a black hole solution.} If we
considered a general perturbation, the departure from the background
solution would be small in the asymptotic regime, so for perturbations
that fall off sufficiently rapidly at large distances, this linearised
analysis gives a relation between the asymptotic behaviour of the
perturbation in the bulk and the stress tensor in the dual field
theory, analogous to that given by the Fefferman-Graham expansion in
the asymptotically AdS case. Note however that for $z \geq 2$, the
falloff of some parts of the bulk perturbation is too slow for this
linearised analysis to be justified, and a full non-linear analysis
will be required even just to relate the asymptotic falloff of the
fields to the boundary stress tensor. This also occurs for the
asymptotically Schr\"odinger case.

We conclude with a summary of our results and a discussion of issues
and directions for further development in section \ref{disc}. In
appendix \ref{deriv}, we calculate the contribution to the stress
tensor for asymptotically Lifshitz spacetimes from counterterms
involving derivatives of the boundary fields. In appendix \ref{Euclidean},
as yet another consistency check, we show that our definition of the
energy density following from the stress tensor complex agrees with
the thermodynamic energy which would be obtained from the Euclidean
action for static asymptotically Lifshitz black holes.

\section{Action for Lifshitz case}
\label{actionsec}

In \cite{klm}, it was proposed that a holographic dual to a theory
with the anisotropic scaling symmetry \eqref{e1} and no boost symmetry
could be obtained by considering the metric\footnote{We always use
  coordinates such that the boundary is at $r = \infty$.}
\begin{equation} \label{lifmet}
ds^2 = - r^{2z} dt^2 + r^2 (dx^2 + dy^2) + \frac{dr^2}{r^2},
\end{equation}
where the scaling symmetry is realised as an isometry: $t \to
\lambda^z t, x^i \to \lambda x^i, r \to \lambda^{-1} r$. This was
realised in \cite{klm} as a solution of a theory with two $p$-form
gauge fields, with a Chern-Simons coupling between the two gauge
fields. In \cite{taylor}, it was observed that one could construct a
simpler theory with the metric and a massive vector field by
integrating out one of the $p$-form gauge fields. We will consider
this case, as it usefully restricts the form of the counter-terms we
can consider in constructing an action principle.
The equations of motion for this theory are
\begin{equation} \label{Eeqn}
R_{\mu\nu} = \Lambda g_{\mu\nu} + \frac{1}{2} F_{\mu\lambda} F_\nu^{\ \lambda} -
\frac{1}{8} F_{\lambda\rho} F^{\lambda \rho} g_{\mu\nu}  +
\frac{1}{2} m^2 A_\mu A_\nu
\end{equation}
and
\begin{equation} \label{Meqn}
\nabla_\mu F^{\mu\nu} = m^2 A^\nu.
\end{equation}
If we choose $\Lambda = -\frac{1}{2}(z^2+z+4)$ and $m^2=2z$, this
theory has a solution
\begin{equation} \label{lmet}
ds^2 = - r^{2z} dt^2 + r^2 (dx^2 + dy^2) + \frac{dr^2}{r^2}, \quad A =
\alpha r^z dt,\quad \alpha^2 = \frac{2(z-1)}{z}.
\end{equation}
It is straightforward to extend
the analysis to a general number of spatial dimensions, but we will
focus on the case of two spatial dimensions for simplicity. We keep
$z$ general; in the linearised analysis we will find that $z=2$ is a
special case, where some aspects of the analysis need separate
treatment.

We want to define an action for this theory which satisfies $\delta S
=0$ with appropriate boundary conditions by adding appropriate local
counter-terms. To preserve the diffeomorphism invariance of the
  action, these counter-terms should be covariant in the boundary
  fields. We consider
\begin{eqnarray}
S =S_{bulk}+S_{bdy}&=&\frac{1}{16 \pi G_4} \int d^4x \sqrt{-g} (R - 2\Lambda -
\frac{1}{4} F_{\mu\nu} F^{\mu\nu} - \frac{1}{2} m^2 A_\mu A^\mu) \\ &&+
\frac{1}{16 \pi G_4} \int d^3 \xi \sqrt{-h} (2K - 4 + f(A_\alpha
A^\alpha)) + S_{deriv} \nonumber ,
\end{eqnarray}
where $\xi^\alpha$ are coordinates on the boundary at some constant
$r$, $h_{\alpha\beta}$ is the induced metric, and
$K_{\alpha\beta}=\nabla_{(\alpha}n_{\beta)}$ is the extrinsic
curvature of the boundary, where the unit vector $n^{\mu}$ is
orthogonal to the boundary and outward-directed. $S_{deriv}$ is a
collection of terms involving derivatives of the boundary fields,
which could involve both the curvature tensor constructed from the
boundary metric and covariant derivatives of $A_\alpha$.  Since the
boundary fields are constants for \eqref{lmet}, as the boundary is
flat, this part of the action will not contribute to the on-shell
value of the action for the pure Lifshitz solution or its first
variation around the Lifshitz background. We can therefore ignore it
for this section, but it can play a role when we come to consider
general asymptotically Lifshitz spacetimes.  The only scalar we can
build from $A$ on the boundary is $A_\alpha A^\alpha$, as $A$ is
constant along the boundary. For the Lifshitz spacetime, $A_\alpha
A^\alpha =-\alpha^2$ is constant, so any function of this scalar will
contribute to the action at the same order in $r$ at large $r$, which
is why we consider an arbitrary function $f(A^\alpha A_\alpha)$ in our
boundary term. For simplicity, we will choose units such that $16
  \pi G_4 = 1$ henceforth.

The variation of the action about a solution of the equations of
motion is just a boundary term,
\begin{eqnarray}
\delta S &=&  \int d^3 \xi \sqrt{-h} \left[
  (\pi_{\alpha \beta} + 2 h_{\alpha \beta}) \delta h^{\alpha \beta} -
  n^\mu F_{\mu\nu} \delta A^\nu \right. \\ && \left. + f'(A_\alpha A^\alpha) (2 A_\alpha
  \delta A^\alpha + A_\alpha A_\beta \delta h^{\alpha \beta})  -
  \frac{1}{2} f(A_\alpha A^\alpha) h_{\alpha \beta} \delta h^{\alpha
  \beta} \right], \nonumber
\end{eqnarray}
where $\pi_{\alpha \beta} = K_{\alpha \beta} - K h_{\alpha\beta}$. For
the Lifshitz spacetime \eqref{lmet}, $\pi_{tt} + 2h_{tt} = 0$,
$\pi_{ij} + 2 h_{ij} = (1-z) r^2 \delta_{ij}$, and $n^\mu F_{\mu \nu}
\delta A^\nu = z \alpha r^z \delta A^t$. Therefore, there are
variations involving $\delta h^{ij}$ and $\delta A^t$ that we need to
cancel. However, the variation involving $\delta h^{tt}$ has already
canceled. To avoid generating a new one from the terms involving
$f(A_\alpha A^\alpha)$, we must have $f(A_\alpha A^\alpha) = \beta
\sqrt{-A_\alpha A^\alpha}$ (so that the $\sqrt{h^{tt}}$ in this
cancels the $\sqrt{h_{tt}}$ in the overall $\sqrt{-h}$ to give us a
term which does not involve $h_{tt}$). Requiring the cancellation of
the other terms determines $\beta = -z \alpha$.  The action is thus
\begin{eqnarray} \label{action}
S &=&  \int d^4x \sqrt{-g} (R - 2\Lambda -
\frac{1}{4} F_{\mu\nu} F^{\mu\nu} - \frac{1}{2} m^2 A_\mu A^\mu) \\ &&+
 \int d^3 \xi \sqrt{-h} (2K - 4 - z\alpha \sqrt{ -
  A_\alpha A^\alpha} )+ S_{deriv}. \nonumber
\end{eqnarray}
It is remarkable that fixing a single coefficient suffices to cancel
both the divergences associated with $\delta h^{ij}$ and $\delta
A^t$. Let us define
\begin{equation} \label{sab}
s_{\alpha\beta} = \sqrt{-h} \left[ (\pi_{\alpha \beta} + 2 h_{\alpha
    \beta}) +\frac{z \alpha}{2}
  (-A_\alpha A^\alpha)^{-1/2} (A_\alpha A_\beta - A_\gamma A^\gamma
  h_{\alpha \beta}) \right] + s_{\alpha\beta}^{deriv},
\end{equation}
\begin{equation} \label{sa}
s_\alpha = -\sqrt{-h} (n^\mu F_{\mu\alpha} -z \alpha
(-A_\alpha A^\alpha)^{-1/2} A_\alpha) + s_\alpha^{deriv}.
\end{equation}
Then the general variation of the action is
\begin{equation}
\delta S =  \int d^3 \xi (s_{\alpha \beta} \delta
h^{\alpha\beta} + s_\alpha \delta A^\alpha).
\end{equation}
In the background \eqref{lmet}, we have $s_{\alpha\beta} = 0$,
$s_\alpha=0$ due to cancellations between the different terms, and
this action satisfies $\delta S = 0$ for arbitrary variations around
\eqref{lmet}. It also has $S=0$ for \eqref{lmet}.

Thus, we have a finite on-shell action which defines a good
variational principle for our background spacetime. Note that an
asset of working with the massive vector theory is that the form of
the possible local counterterms is tightly constrained; with the
original theory of \cite{klm}, we could build several different
scalars from the two $p$-forms, and it would not be so obvious what a
convenient form for the action is. In section~\ref{pert}, we will show
that the action is finite on-shell and gives a well-defined
variational principle for a class of asymptotically Lifshitz
spacetimes.  Before doing so, however, we want to address the
calculation of the stress tensor from the action in asymptotically
Lifshitz and asymptotically Schr\"odinger spacetimes.

\section{Stress tensor}
\label{stress}

A core element of the holographic renormalization programme in the
gauge-gravity duality for relativistic field theories is that given a
well-defined action principle, we can use it to define a boundary
stress tensor as the variation of the action with respect to the
boundary metric~\cite{Henningson:1998gx,Balasubramanian:1999re}. The
resulting stress tensor has been shown to define conserved charges
which generate the asymptotic symmetries of the geometry in very
general circumstances \cite{him}. This stress tensor carries important
physical information about the dual field theory. In this section, we
want to discuss the calculation of such a boundary stress tensor from
a bulk action principle in the non-relativistic case. We will focus
explicitly on the Lifshitz example in this section, as its treatment
is simpler, but similar ideas apply to asymptotically Schr\"odinger
spacetimes, which we consider in the next section.

For asymptotically Lifshitz spacetimes, the dual field theory is
non-relativistic, so it will not have a covariant relativistic stress
tensor, but we would still expect it to have a stress tensor complex,
consisting of the energy density ${\mathcal E}$, energy flux
${\mathcal E}_i$, momentum density ${\mathcal P}_i$ and spatial stress
tensor $\Pi_{ij}$, satisfying the conservation equations
\begin{equation} \label{cons}
\partial_t {\mathcal E} + \partial_i {\mathcal E}^i = 0, \quad
\partial_t {\mathcal P}_j + \partial_i \Pi^i_{\ j} = 0.
\end{equation}
We would like to derive such a stress tensor complex by considering
some appropriate variations of the action principle we introduced in
the previous section. Since the boundary theory is non-relativistic,
the boundary data does not include a non-degenerate metric; the
nonuniform $r$-dependence of the metric in the bulk along the boundary
directions leads to a degenerate boundary metric. It is therefore not
a priori obvious how to define the stress tensor complex. In this
section we will follow the relativistic analysis as closely as
possible; we postpone discussion of the appropriateness of this
approach from the boundary theory point of view to the conclusions.

Since the background \eqref{lmet} involves a vector field, we will
need to consider how this effects the definition of the stress
tensor. This issue was considered in the relativistic case
in~\cite{him}, where it was argued that the appropriate definition of
the stress tensor in the presence of tensor fields was to consider the
variation of a boundary frame field $\hat{e}_\alpha^{(A)}$, holding
the tangent space components $\phi^{[i]}_{AB\ldots}$ of the other
fields fixed where $A$,~$B$,$\cdots$ denote tangent space directions
and $i$ denotes matter species. This was shown to provide a stress
tensor whose integrals give the conserved charges generating
asymptotic symmetries and which is conserved up to terms involving
derivatives of the other fields \cite{him}.

To be more specific, if we considered a background with a massive
vector field which was dual to a relativistic field theory, we should
hold the components $A_A$ of the vector with tangent space indices
fixed. We would then write the general variation of the action as
\begin{equation}
\delta S = \int \hat{\epsilon} (T^\alpha_A \delta \hat{e}_\alpha^{(A)} + s_A
\delta A^A),
\end{equation}
where $\hat{e}_\alpha^{(A)}$ is a boundary frame field defining the
boundary metric, and $\hat{\epsilon}$ is the associated volume form on
the boundary. That is, $\hat{e}_\alpha^{(A)}$ are the components of the
frame along the boundary directions, rescaled by an appropriate power
of $r$ such that $\hat{e}_\alpha^{(A)}$ have finite limits as $r \to
\infty$. In an asymptotically AdS spacetime, the choice of
$\hat{e}_\alpha^{(A)}$ corresponds to the choice of the boundary metric
$g_{(0)}$ appearing in the expansion of the asymptotic geometry in
Fefferman-Graham coordinates,
\begin{equation}
ds^2_{AdS} = \frac{dr^2}{r^2} + r^{2} [g_{(0) \alpha \beta} +
{\mathcal O}(r^{-2}) ] dx^\alpha dx^\beta,
\end{equation}
and the bulk frame fields are related to the boundary frame fields by
$e^{(A)} = r \hat{e}^{(A)}$, $e^{(r)} = \frac{dr}{r}$.  The stress tensor
$T^\alpha_A$ was shown in \cite{him} to be conserved up to terms
involving the variation of the matter fields,
\begin{equation} \label{relcons}
D_\alpha T^{\alpha}_{\ \beta} = s_A \partial_\beta A^A,
\end{equation}
where $D_\alpha$ is the covariant derivative on the boundary defined
by requiring $D_\alpha \hat{e}_\beta^{(B)} = 0$.  In the
asymptotically AdS case, the key advantage of the prescription of
\cite{him} is that it gives a stress tensor which is conserved in this
sense. If we considered the stress tensor as defined by considering
the variation of the metric holding the spacetime components of the
matter fields fixed, we would obtain a finite result, but there would
be additional terms on the right-hand side of this conservation
equation, and as a result, the stress tensor would not in general give
rise to the correct conserved charges (although the difference is
unimportant in many common examples). In the non-relativistic cases,
as we will see below, this distinction is much more important, and we
must follow the prescription of \cite{him} to obtain finite results
for the stress tensor complex.

We want to apply a similar prescription to our non-relativistic
cases. In asymptotically Lifshitz spacetimes, because of the different
scaling of the time and space directions, there is no non-degenerate
boundary metric that we can associate with the boundary at $r =
\infty$ in our spacetime.  However, when we calculate the variation in
\eqref{action}, we first cut off the spacetime at some finite radius
$r$, and then consider the limit as $r \to \infty$. At finite $r$,
there is a well-defined boundary metric. We could rescale the bulk
metric by $r^2$ so that the spatial parts have a well-defined large
$r$ limit; the additional factor of $r^{2(z-1)}$ multiplying $dt^2$
can then be thought of as a radius-dependent speed of light, so that
the limit $r \to \infty$ corresponds to taking the speed of light to
infinity in the boundary theory. In the non-relativistic limit of a
relativistic field theory, we can recover both of the conservation
equations \eqref{cons} from the conservation of the relativistic
stress tensor. If we take this point of view, then we should expect to
be able to define the non-relativistic stress tensor complex following
essentially the same recipe as in the relativistic
case.\footnote{Working on a finite cutoff surface in this way is also
  similar in spirit to the analysis of holographic renormalisation for
  asymptotically flat spaces in \cite{Mann:2005yr}.}

For the Lifshitz case, we assume that we have a bulk orthonormal frame
with components
\begin{equation}
  e^{(0)} = r^z \hat{e}^{(0)}, \quad e^{(i)} = r
  \hat{e}^{(i)}, \quad e^{(3)} = \frac{dr}{r},
\end{equation}
and a massive vector $A_M$. From the heuristic point of view above,
the different scaling in $e^{(0)}$ compared to $e^{(i)}$ corresponds
to a scaling by the radius-dependent speed of light on the surface at
constant $r$. We will use indices $M = 0,1,2,3$ to denote frame
components and $\mu =t,x,y,r$ to denote spacetime components. The
spacetime will asymptotically approach the pure Lifshitz solution
\eqref{lmet} if $\hat{e}^{(0)} \to dt$, $\hat{e}^{(i)} \to dx^i$ and
$A^M \to \alpha \delta^M_0$ as $r \to \infty$.\footnote{This is a
  necessary condition; we will give a more precise definition of
  asymptotically Lifshitz boundary conditions for more general
  boundary data later.}

We therefore construct the stress tensor complex for the
non-relativistic theories by regarding $\hat{e}^{(0)}$, $\hat{e}^{(i)}$ and
$A^M$ (more accurately, their limits as $r \to \infty$) as the
boundary data, and defining
\begin{equation} \label{nrels}
\delta S = \int \hat{\epsilon} [-{\mathcal E} \delta \hat{e}_t^{(0)} -
{\mathcal E}^i \delta \hat{e}_i^{(0)} + {\mathcal P}_i \delta \hat{e}^{(i)}_t
+ \Pi_i^j \delta \hat{e}^{(i)}_j + s_A \delta A^A].
\end{equation}
As in the relativistic case, we expect that the energy density, energy
flux, momentum density and spatial stress tensor so defined will
satisfy the conservation equations \eqref{cons} up to terms involving
the variation of the massive vector field. The treatment of the matter
fields, holding the components with tangent space indices fixed, turns
out to be crucial to obtain finite results for the stress tensor.

If the boundary data are taken to be $\hat{e}^{(0)} \to dt$, $\hat{e}^{(i)}
\to dx^i$, then $\hat \epsilon$ is just the flat volume form $d^3
\xi$, and we can rewrite the above definitions in terms of the
coefficients $s_{\alpha \beta}$ and $s_\alpha$ that we used to write
the general metric variation in \eqref{action}:
\begin{equation}
{\mathcal E} = 2 s^t_{\ t} - s^t A_t, \quad
{\mathcal E}^i = 2s^i_{\ t} - s^i A_t,
\end{equation}
and
\begin{equation}
{\mathcal P}_i = -2 s^t_{\ i} + s^t A_i  \quad \Pi_{i}^j =-2 s^j_{\
  i} + s^j A_i,
\end{equation}
where we have multiplied through by factors of the frame fields to
simplify the form of these expressions, taking advantage of the fact
that the frame fields each have a single component to leading order in
the large $r$ limit, so all indices are now spacetime indices.  Note
that when $z=1$, $\alpha=0$ and these definitions reduce to the
familiar AdS rules.

Normally, to obtain finite quantities in the non-relativistic limit of
the relativistic stress tensor, we need to eliminate divergent
contributions to the energy density and energy flux coming from the
rest mass of the particles (see e.g. \cite{ll} chapter 15). However,
these Lifshitz theories do not have Galilean boost invariance, and
hence do not conserve particle number. We will find below that the
above definitions give a finite result for the energy density,
indicating that there is no divergent contribution from rest mass that
we need to eliminate.

\section{Schr\"odinger spacetimes}
\label{schr}

Another example of non-relativistic holography is the Schr\"odinger
spacetime \cite{son2008,Balasubramanian:2008dm},
\begin{equation}
ds^2 = -r^4 (dx^+)^2 + r^2 (-2 dx^+ dx^- + d\mathbf{x}^2) +
\frac{dr^2}{r^2}.
\end{equation}
This solution has the Schr\"odinger symmetry group as its isometries
(including in particular the anisotropic scaling symmetry \eqref{e1}
with $z=2$, when we identify $t$ there with $x^+$). It was shown
  in \cite{Adams:2008zk,SchaferNameki:2009xr} that this symmetry group
  essentially uniquely determines this form for the metric.

A simple action which has a solution with this metric is \cite{heat}
\begin{eqnarray} \label{scact}
S &=& \frac{1}{16 \pi G_5} \int d^5 x \sqrt{-g} \left[ R -
  \frac{4}{3} \partial_\mu \phi \partial^\mu \phi - \frac{1}{4}
  e^{-8\phi/3} F_{\mu\nu} F^{\mu\nu} - 4 A_\mu A^\mu - V(\phi) \right]
\\ && +\frac{1}{16 \pi G_5} \int d^4 \xi \sqrt{-h} \left[ 2K-6+ (1+c_4
  \phi) A_\mu
  A^\mu + c_5 (A_\mu A^\mu)^2 + (2c_4 - 4c_5
  +3) \phi^2 \right],  \nonumber
\end{eqnarray}
which gives
\begin{equation}
\delta S = \frac{1}{16 \pi G_5} \int d^4 \xi (s_{\alpha
  \beta} \delta h^{\alpha \beta} + s_\alpha \delta A^\alpha + s_\phi
\delta \phi),
\end{equation}
with
\begin{eqnarray}
s_{\alpha \beta} &=& \sqrt{-h} [\pi_{\alpha\beta} + 3 h_{\alpha \beta} + (1+ c_4
\phi) (A_\alpha A_\beta - \frac{1}{2} A_\gamma A^\gamma h_{\alpha
  \beta}) \\ && +c_5 A_\delta A^\delta (2 A_\alpha A_\beta - \frac{1}{2}
A_\gamma A^\gamma h_{\alpha \beta}) - \frac{1}{2} (2c_4 - 4 c_5 + 3
) \phi^2 h_{\alpha \beta}], \nonumber
\end{eqnarray}
\begin{equation}
s_\alpha = \sqrt{-h}(- n^\mu F_{\mu \alpha} e^{-8\phi/3} + 2(1+c_4 \phi +2 c_5
A_\gamma A^\gamma) A_\alpha),
\end{equation}
and
\begin{equation}
s_\phi = \sqrt{-h} (-\frac{8}{3} n^\mu \partial_\mu \phi + c_4
A_\alpha A^\alpha + 2(2c_4-4c_5+3) \phi).
\end{equation}

This has an asymptotically Schr\"odinger black hole solution
\cite{heat,mmt,abm}. The metric is
\begin{eqnarray}
ds_E^2 &=& r^2\, k(r)^{-\frac{2}{3}}\left(\left[\frac{1-f(r)}{4\beta^2} -
    r^2\,f(r)\right] \, (dx^+)^2 + \frac{\beta^2 r_+^4}{r^4} \, (dx^-)^2 - \left[1+f(r)\right]\,dx^+\,dx^- \right) \nonumber \\
&& \quad+\;\;  k(r)^{\frac{1}{3}}\, \left(r^2 d {\bf x}^2 +  \frac{dr^2}{r^2\, f(r)} \right),
\label{5dbh}
\end{eqnarray}
with the massive vector and scalar
\begin{eqnarray}
A &=& \frac{r^2 }{k(r)} \, \left( \frac{1+f(r)}{2}\, dx^+ - \frac{\beta^2 r_+^4}{r^4}\, dx^-\right), \nonumber \\
 e^\phi &=& \frac{1}{\sqrt{k(r)}} \ .
\label{5doth}
\end{eqnarray}
This solution is obtained by applying a solution-generating
transformation (the TsT transformation) to an asymptotically AdS
vacuum black hole solution.  In \cite{heat}, it was shown that the
action \eqref{scact} is finite and satisfies $\delta S=0$ for this
black hole solution. However, some of the coefficients $s_{\alpha\beta}$ in
the variation diverge, so a naive attempt to define the stress tensor
will fail \cite{heat,Yamada:2008if}.

As in the asymptotically Lifshitz case, there is no non-degenerate
boundary metric for the asymptotically Schr\"odinger
spacetimes. However, as before, there is a non-degenerate metric on
the surfaces of finite $r$, which degenerates in the limit as $r \to
\infty$. We therefore define a stress tensor complex for these
spacetimes by adapting the relativistic prescription in \cite{him}. In
this case, the non-relativistic theory is meant to be obtained from
the relativistic theory by light-cone reduction, with the momentum
along the light cone direction interpreted as the conserved mass
density $\rho$, which satisfies a conservation equation involving the
mass flux $\rho^i$. The combination which appears as the coefficient
of $\delta e^{(A)}_\alpha$ in $\delta S$ is again $-2 s^\alpha_{\ A} +
s^\alpha A_A= (-2 s^\alpha_{\ \beta} + s^\alpha A_\beta) e^\beta_{\
  {(A)}}$. There is no obvious convenient choice of orthonormal
frame. We therefore identify the components of the stress tensor
complex in this case as
\begin{equation} \label{securr}
{\mathcal E} = 2s^+_{\ +} -s^+ A_+, \quad {\mathcal E}^i = 2 s^i_{\ +} -
s^i A_+,
\end{equation}
\begin{equation} \label{smcurr}
{\mathcal P}_j = -2s^+_{\ j} + s^+ A_j, \quad \Pi^i_{\ j} = -2 s^i_{\ j} + s^i A_j,
\end{equation}
and
\begin{equation} \label{spcurr}
\rho = -2 s^+_{\ -} + s^+ A_-, \quad \rho^i = -2 s^i_{\ -} + s^i A_-,
\end{equation}
where all the indices are again spacetime indices, and we have
  set $16 \pi G_5 = 1$.

For the black hole solution \eqref{5dbh}, all of the vector components
of the stress tensor complex vanish, and we find
\begin{equation}
{\mathcal E} = r_+^4, \quad \Pi_{xx} = \Pi_{yy} = r_+^4, \quad \rho
= 2\beta^2 r_+^4,
\end{equation}
in agreement with previous results obtained by different methods
\cite{heat,hydro}.  Note that because of the slow falloff
relative to the background, there is a potential finite $\beta^4
r_+^8$ term in ${\mathcal E}$, that is, a piece which comes from terms
quadratic in the departure from the background. It is a non-trivial
test of our definition of the stress tensor that this term vanishes.

For these asymptotically Schr\"odinger spacetimes, it was proposed in
\cite{mmt} that the stress tensor could be obtained by taking the
stress tensor for the corresponding asymptotically AdS spacetime and
taking the light cone reduction of it. This idea was applied to the
study of the hydrodynamics for the non-relativistic theories with
Schr\"odinger symmetry in \cite{hydro}. It is therefore
important for us to compare this approach to our new proposal.

These two approaches a priori look quite different; one reason why we
might nevertheless expect agreement is that the stress tensor was
shown in \cite{him} to give the conserved charges associated with the
asymptotic symmetries of the spacetime. In the Schr\"odinger case, the
action of symmetries like time translation will commute with the TsT
transformation, so we can perform a time translation by transforming
to the asymptotically AdS spacetime, performing a time translation
there, and transforming back to the asymptotically Schr\"odinger
spacetime. Thus, the conserved charge obtained from the stress tensor
of \cite{mmt}, which generates time translation in the asymptotically
AdS spacetime, is naturally identified with the conserved charge which
generates time translation in the asymptotically Schr\"odinger
spacetime. This provides some physical motivation for agreement of the
two stress tensors.

For simplicity, let us consider an asymptotically Schr\"odinger
spacetime which is obtained by a TsT transformation from a vacuum
asymptotically AdS spacetime. This does not give the most general
asymptotically Schr\"odinger spacetime (which would require us to
consider an asymptotically AdS spacetime with non-zero scalar and
massive vector fields in the bulk), but restricting consideration to
this case leads to much simpler expressions, and includes all the
examples that have been explicitly considered so far in the
literature. If we start with a vacuum asymptotically AdS solution with
metric
\begin{equation}
ds^2_{AdS} = \bar{g}_{\mu\nu} dx^\mu dx^\nu = \bar{g}_{\alpha \beta}
dx^\alpha dx^\beta + \frac{dr^2}{r^2},
\end{equation}
and we assume that the metric is independent of a coordinate $x^-$
which becomes null at large distances, then by applying a TsT
transformation we will obtain an asymptotically Schr\"odinger solution
with scalar field
\begin{equation} \label{lphir}
e^{-2\phi} = 1 +
\bar{g}_{--},
\end{equation}
massive vector field
\begin{equation}
A_\mu = e^{2 \phi} \bar{g}_{\mu-}, \quad
\end{equation}
and metric
\begin{equation} \label{lmetr}
g_{\mu\nu} = e^{-2 \phi/3} \bar{g}_{\mu\nu} - e^{4\phi/3}
\bar{g}_{\mu-} \bar{g}_{\nu-},
\end{equation}
which implies the inverse metric is
\begin{equation}
g^{\mu\nu} = e^{2\phi/3} (\bar{g}^{\mu\nu} + \delta^\mu_-
\delta^\nu_-).
\end{equation}

Our definition of the non-relativistic stress tensor complex for the
asymptotically Schr\"odinger spacetime corresponds to considering the
light cone reduction of a ``stress tensor''
\begin{equation}
T^\alpha_{\ \beta} = s^\alpha_{\ \beta} - \frac{1}{2} s^\alpha A_\beta =
 \sqrt{-h} h^{\alpha \gamma} \tau_{\gamma \beta}
\end{equation}
where
\begin{eqnarray}
\tau_{\gamma\beta} &=& \pi_{\gamma \beta} + \frac{1}{2} e^{-8 \phi/3} n^\mu F_{\mu\gamma}
  A_\beta  \\ &&- \frac{1}{2} \left(-6 + (1+c_4 \phi) A_\delta A^\delta + c_5
  (A_\delta A^\delta)^2 + (2c_4 - 4c_5 + 3) \phi^2 \right) h_{\gamma
\beta},\nonumber
\end{eqnarray}
whereas \cite{mmt} would consider the stress tensor for the
asymptotically AdS spacetime, which is simply
\begin{equation}
  \bar{T}^\alpha_{\ \beta} = \sqrt{-\bar{h}} \bar{h}^{\alpha \gamma} [ \bar{\pi}_{\gamma \beta} +
  3 \bar{h}_{\gamma \beta} ].
\end{equation}
To compare these two, let's rewrite our stress tensor using the
expression for the Schr\"odinger fields in terms of the AdS
metric. The unit normal in the asymptotically Schr\"odinger geometry
is $n^\mu = r e^{\phi/3} \delta^\mu_r$, so
\begin{eqnarray}
K_{\alpha \beta} &=& \frac{1}{2} r e^{\phi/3} \left(
  -\frac{2}{3} \partial_r \phi e^{-2\phi/3} \bar{h}_{\alpha \beta} +
  e^{-2\phi/3} \partial_r \bar{h}_{\alpha \beta}  \right. \\ && \left. \nonumber -
  \frac{4}{3} \partial_r \phi e^{4 \phi/3} \bar{h}_{\alpha-}
  \bar{h}_{\beta-}-  e^{4 \phi/3} \partial_r \bar{h}_{\alpha-}
  \bar{h}_{\beta-}-  e^{4 \phi/3} \bar{h}_{\alpha-}
  \partial_r  \bar{h}_{\beta-} \right),
\end{eqnarray}
which gives
\begin{equation}
\pi_{\alpha \beta} = e^{-\phi/3} \bar{\pi}_{\alpha \beta} +
\frac{1}{2} r e^{5 \phi/3} \left( -2 \partial_r \phi  \bar{h}_{\alpha-}
  \bar{h}_{\beta-} -  \partial_r   \bar{h}_{\alpha-}
  \bar{h}_{\beta-} -   \bar{h}_{\alpha-}
  \partial_r \bar{h}_{\beta-}+ \bar{h}^{\gamma \delta} \partial_r
  \bar{h}_{\gamma \delta} \bar{h}_{\alpha-}
  \bar{h}_{\beta-} \right),
\end{equation}
and we have
\begin{equation}
\frac{1}{2} n^\mu e^{-8 \phi/3} F_{\mu \alpha} A_\beta  = \frac{1}{2} r e^{5 \phi/3} \left( 2 \partial_r \phi  \bar{h}_{\alpha-}
  \bar{h}_{\beta-} +  \partial_r   \bar{h}_{\alpha-}
  \bar{h}_{\beta-} \right).
\end{equation}
Thus,
\begin{eqnarray} \label{treln}
\tau_{\alpha \beta} &=&  e^{-\phi/3} \bar{\pi}_{\alpha \beta} +
\frac{1}{2} r e^{5 \phi/3} (-   \bar{h}_{\alpha-}
  \partial_r \bar{h}_{\beta-}+ \bar{h}^{\gamma \delta} \partial_r
  \bar{h}_{\gamma \delta} \bar{h}_{\alpha-}
  \bar{h}_{\beta-} ) \\ && \nonumber -
\frac{1}{2} \left(-6 + (1+c_4 \phi) A_\alpha A^\alpha + c_5
  (A_\alpha A^\alpha)^2 + (2c_4 - 4c_5 + 3) \phi^2 \right)
h_{\alpha \beta}  \\ &=& \nonumber  e^{-\phi/3} ( \bar{\pi}_{\alpha
  \beta} + 3 \bar{h}_{\alpha \beta} ) - e^{5 \phi/3} \bar{h}_{\alpha
  -} (\bar{\pi}_{\beta -} + 3 \bar{h}_{\beta -})  \\ &&- \nonumber
 \frac{1}{2}  \left(6 e^{\phi/3} -6 + (1+c_4 \phi) A_\alpha A^\alpha + c_5
  (A_\alpha A^\alpha)^2 + (2c_4 - 4c_5 + 3) \phi^2 \right)
h_{\alpha \beta}.
\end{eqnarray}
The two expressions are thus clearly not manifestly the same. However,
to compare them we should consider the behaviour at large
$r$.

The asymptotically AdS solution has
\begin{equation}
\bar{h}_{\alpha \beta} = r^2 \eta_{\alpha \beta} + \frac{1}{r^2}
\bar{h}^{(1)}_{\alpha \beta}.
\end{equation}
This implies that $h^{\alpha \beta} \sim r^{-2}$ except for $h^{--}
\sim r^0$, and $\sqrt{-h} \sim r^4$, so for $\alpha \neq -$, finite
contributions to $T^\alpha_{\ \beta}$ come from terms in $\tau_{\gamma
  \beta}$ which go like $r^{-2}$, and we can neglect any contribution
which falls off more rapidly. We have $\bar{\pi}_{\alpha\beta} + 3
\bar{h}_{\alpha \beta} \sim r^{-2}$, so the first term in
\eqref{treln} gives a finite contribution. For $\alpha \neq +$,
$\bar{h}_{\alpha -} \sim r^{-2}$, so the second term can be
neglected. For $\alpha = +$, however, the second term gives a
potentially divergent contribution to the stress tensor. To calculate
the last term in \eqref{treln}, it is useful to note that
\begin{equation}
A_\alpha A^\alpha = e^{8 \phi/3} \bar{g}_{--}.
\end{equation}
We then find
\begin{eqnarray}
G &\equiv& 6 e^{\phi/3} -6 + (1+c_4 \phi) A_\alpha A^\alpha + c_5 (A_\alpha
A^\alpha)^2 + (2c_4 - 4c_5 + 3) \phi^2 \\ \nonumber &=& \left( \frac{83}{216} +
  \frac{5}{12} (c_4-4c_5) \right) \bar{g}_{--}^3 + \ldots \sim \frac{1}{r^6},
\end{eqnarray}
where the dots denote terms of higher order in a Taylor expansion in
$\bar{g}_{--}$. The last term hence can be neglected, except when $\alpha
= \beta = +$ (as $h_{++} \sim r^4$ at large $r$).

Thus, for all the components where $\alpha \neq +$,
\begin{equation}
\tau_{\alpha \beta} = \bar{\pi}_{\alpha \beta} + 3
\bar{h}_{\alpha \beta} + {\mathcal O}(r^{-4}).
\end{equation}
For $\alpha = +$,
$\beta \neq +$,
\begin{equation} \label{tpb}
\tau_{+\beta} = - e^{5 \phi/3} \bar{h}_{+-} (
\bar{\pi}_{\beta -} +3 \bar{h}_{\beta -}) + e^{-\phi/3}
(\bar{\pi}_{\beta+} + 3 \bar{h}_{\beta +}) + {\mathcal O}(r^{-6}),
\end{equation}
where the first term is order $r^0$, and the second term is order $r^{-2}$. For
$\alpha = +$, $\beta = +$,  there is an order $r^{-2}$ term from
the last term in \eqref{treln}, so
\begin{equation}
\tau_{++} = - e^{5 \phi/3} \bar{h}_{+-} (
\bar{\pi}_{+ -} +3 \bar{h}_{+ -}) + {\mathcal O}(r^{-2}).
\end{equation}

Let us now consider the implications of this asymptotic behaviour for
our non-relativistic stress tensor complex. The non-relativistic
  stress tensor complex defined in (\ref{securr},\ref{smcurr},\ref{spcurr}) is
  constructed from the components $T^\alpha_{\ \beta}$ with $\alpha
  \neq -$, so we are mainly interested in these. For $\alpha = i$,
\begin{equation}
T^i_{\ \beta} = \sqrt{-h} h^{i \gamma} \tau_{\gamma \beta} =
e^{2\phi/3} r^2
\tau_{i \beta} + {\mathcal O}(r^{-2}) = \bar{T}^i_{\ \beta} +
{\mathcal O}(r^{-2}).
\end{equation}
Similarly, for $\alpha = +$,
\begin{equation}
T^+_{\ \beta} = \sqrt{-h} h^{+ \gamma} \tau_{\gamma \beta} = e^{2\phi/3} r^2
\tau_{- \beta} + {\mathcal O}(r^{-2}) = \bar{T}^+_{\ \beta} +
{\mathcal O}(r^{-2}).
\end{equation}
Thus, for the components that contribute to our definition of the
non-relativistic stress tensor complex, we find precise agreement with
the definition of \cite{mmt}. Note in particular that $\tau_{+ \beta}$
will not affect these contributions, as $h^{+i}, h^{++} \sim r^{-6}$.
Thus, our definition of the non-relativistic stress tensor complex and
the definition proposed in \cite{mmt} will agree on asymptotically
Schr\"odinger solutions which are obtained by TsT transformation from
a vacuum asymptotically AdS solution.

It is also interesting to consider what happens for the remaining
components of the stress tensor, those with $\alpha = -$. We have
\begin{eqnarray}
T^-_{\ \beta} &=& \sqrt{-h} h^{- \gamma} \tau_{\gamma \beta} =
\sqrt{-h} e^{2\phi/3} [ \bar{h}^{- \gamma} \tau_{\gamma \beta} +
\tau_{-\beta}]  \\ &=& \sqrt{-h} e^{2\phi/3} \bar{h}^{-\gamma}[ e^{-\phi/3}
(\bar{\pi}_{\gamma \beta} + 3 \bar{h}_{\gamma \beta}) - \frac{1}{2}
   G h_{\gamma \beta}] + {\mathcal O}(r^{-2}). \nonumber
\end{eqnarray}
There are two sources of potentially divergent contributions in this
term, coming from the $r^0$ part in $\tau_{+ \beta}$, and the $r^{-2}$
part in $\tau_{- \beta}$. These both involve factors of
$\bar{\pi}_{\beta -} +3 \bar{h}_{\beta -}$, and they cancel exactly to
leave a finite result for this component of the stress tensor. The
term involving $G$ is negligible except for $\gamma = \beta = +$, so
the components $T^-_{\ \beta}$ for $\beta \neq +$ will also agree with
the definition of \cite{mmt}. The component $T^-_{\ +}$, although
finite, will not in general agree with the definition of
\cite{mmt}.\footnote{We could choose the constants $c_4, c_5$ to make
  the $\bar{g}_{--}^3$ contribution to $G$ vanish, and it would then
  agree. However, it is better to use this freedom instead to
  eliminate a divergence in $s_\phi$, as we will shortly describe.}
However, this disagreement does not affect the physics. To make
contact with a non-relativistic theory by light cone reduction, we are
restricting to metrics which are independent of $x^-$. This implies
that the $T^-_{\ \beta}$ drop out of the conservation equations; they
are not part of the conserved currents associated with the restricted
diffeomorphism freedom which preserves the manifest Killing symmetry
along $x^-$. A disagreement in these components thus has no physical
consequences for the non-relativistic dual.

In \cite{heat}, it was shown that the action \eqref{scact} satisfies
$\delta S=0$ for variations around the black hole solution
\eqref{5dbh} satisfying some rather restrictive boundary
conditions. We have shown that the stress tensor complex is finite for
a large family of asymptotically Schr\"odinger solutions. Since our
stress tensor is defined as the variation of the action with respect
to a variation in the asymptotic boundary values of the frame fields,
this implies that $\delta S=0$ for any variation of the frame fields
which does not change the asymptotic boundary values. However, the
coefficients of matter field variations $s_\alpha$, $s_\phi$ will
still diverge for general asymptotically Schr\"odinger solutions (and
in particular for the black hole solution \eqref{5dbh}), so we still
need to impose restrictive boundary conditions on the variations of
the matter fields, as in \cite{heat}. We can make the divergent
contribution to $s_\phi$ vanish by choosing the coefficients in the
action so that $c_4 - 4 c_5 + \frac{17}{3} = 0$, but we are still left
with divergences in $s_\alpha$. A more general understanding of these
asymptotic boundary conditions is an interesting problem for the
future.

\section{Asymptotic perturbation analysis for Lifshitz}
\label{pert}

We want to show that the action \eqref{action} is finite on-shell and
satisfies $\delta S=0$ for a class of asymptotically Lifshitz
spacetimes. Black hole solutions which asymptotically approach
\eqref{lmet} were obtained in \cite{dt,mann}, and we could consider
the behaviour for these backgrounds. However, since the solutions are
only known numerically, a direct analysis of these solutions is
difficult and not very illuminating.\footnote{The only known analytic
  black hole solutions \cite{peet,mann} to (\ref{Eeqn},\ref{Meqn})
  have non-flat boundary.} Instead, it is better to perform a general
analysis of the equations of motion in the asymptotic region. Finding
exact solutions of the equations of motion (\ref{Eeqn},\ref{Meqn})
analytically is difficult. However, if the solution is asymptotically
Lifshitz, it will be a small perturbation of \eqref{lmet} for
sufficiently large $r$. Let us therefore study the solutions of the
linearized equations of motion expanding around \eqref{lmet}. This
calculation will also be useful for obtaining two-point functions on
the background \eqref{lmet}, although we will not investigate this
here. Note that the analysis of the constant scalar perturbations was
also performed in \cite{dt,mann,peet}; perturbative analysis of
related solutions was also performed in \cite{Azeyanagi:2009pr}.

If we write the background as $g_{\mu\nu}$, $A_\mu$ and the
perturbations as $h_{\mu\nu}$, $a_\mu$, then the linearized
equations are\footnote{Note that $h_{\mu\nu}$ denotes
  the perturbation of the metric, and indices are raised and lowered
  with the background metric, so $h^{\mu\nu}$ is the perturbation of
  the metric with the indices raised, not the perturbation of the
  inverse metric. This differs from the convention in the discussion
  of the variation of the action, where $\delta h^{\mu\nu}$ is the
  variation of the inverse metric.}
\begin{equation}
\nabla_\mu f^{\mu\nu} - \nabla_\mu (h^{\mu\lambda} F_\lambda^{\ \nu})
- \nabla_\mu h^{\beta \nu} F^\mu_{\ \beta}
+ \frac{1}{2} \nabla_\lambda h F^{\lambda \nu} = m^2 a^\nu
\end{equation}
and
\begin{eqnarray}
R_{\mu\nu}^{(1)} &=& \Lambda h_{\mu\nu} + \frac{1}{2} f_{\mu\lambda} F_\nu^{\ \lambda} +
\frac{1}{2} f_{\nu\lambda} F_\mu^{\ \lambda} - \frac{1}{2}
F_{\mu\lambda} F_{\nu\sigma} h^{\lambda \sigma} -\frac{1}{4}
f_{\lambda \rho} F^{\lambda \rho} g_{\mu\nu} + \frac{1}{4} F_{\lambda
  \rho} F_\sigma^{\ \rho} h^{\lambda \sigma} g_{\mu\nu}  \nonumber \\ &&- \frac{1}{8}
F_{\lambda \rho} F^{\lambda \rho} h_{\mu \nu} + \frac{1}{2} m^2
a_\mu A_\nu + \frac{1}{2} m^2 a_\nu A_\mu,
\end{eqnarray}
where $f_{\mu\nu} = \partial_\mu a_\nu - \partial_\nu a_\mu$ and
\begin{equation}
R_{\mu\nu}^{(1)} = \frac{1}{2} g^{\lambda \sigma} [ \nabla_\lambda
  \nabla_\mu h_{\nu \sigma} + \nabla_\lambda \nabla_\nu h_{\mu\sigma} -
  \nabla_\mu \nabla_\nu h_{\lambda \sigma} - \nabla_\lambda
  \nabla_\sigma h_{\mu\nu} ].
\end{equation}

It is convenient to define
\begin{equation} \label{ans1}
h_{tt} = -r^{2z} \hat{h}_{tt}, \quad h_{ti} = -r^{2z} v_{1i} + r^2
v_{2i}, \quad h_{ij} = r^2 \hat{h}_{ij},
\end{equation}
\begin{equation} \label{ans2}
a_t = \alpha r^z( \hat{a}_t + \frac{1}{2} \hat{h}_{tt}), \quad a_i =
\alpha r^z
v_{1i}, \quad a_r = \alpha \frac{\hat{a}_r}{r}.
\end{equation}
We choose a Gaussian normal gauge, so $h_{r\mu} = 0$. In terms of a
frame field, this definition corresponds to choosing the orthonormal
frame
\begin{equation}
e^{(0)} = r^z \hat{e}^{(0)} = r^z [ (1+ \frac{1}{2} \hat{h}_{tt}) dt +
v_{1i} dx^i ], \quad
e^{(i)} = r \hat{e}^{(i)} = r [v_{2i} dt + (\delta^i_{\ j} +
\frac{1}{2} \hat{h}^i_{\ j} )
dx^j] , \quad e^{(3)} = \frac{dr}{r},
\end{equation}
and the vector field components in the orthonormal frame to be
\begin{equation}
A^M = \alpha(1+\hat{a}_t) \delta^M_{\ 0} + \alpha \hat{a}_r
\delta^M_{\ 3}.
\end{equation}
That is, we are partially fixing the freedom in the choice of frame
(local Lorentz invariance) by choosing the frame vector $e^{(0)}$ to be
parallel to the projection of the vector field $A$ along the boundary
at constant $r$.

For our spacetime to be asymptotically Lifshitz, we will at least
require that the normalised perturbations $\hat{h}_{tt}$, $v_{1i}$,
$v_{2i}$, $\hat{h}_{ij}$, $\hat{a}_t$ and $\hat{a}_r$ all vanish as $r
\to \infty$. In terms of the frame fields, we are saying that a
necessary condition for the spacetime to be asymptotically Lifshitz is
that $\hat{e}^{(0)} \to dt$, $\hat{e}^{(i)} \to dx^i$, $A^M \to \alpha
\delta^M_{\ 0}$ as $r \to \infty$. We will be more precise about our
boundary conditions once we have solved the linearised equations of
motion.

One of our goals is to show that the action \eqref{action} is finite
on-shell. In the linearised analysis, since the background solution
has no vector-like parts in the spatial directions along the boundary
and the action is a scalar, the action to linear order will only
involve the scalar parts of the linearised perturbations. Furthermore,
the integration over the boundary directions makes the value of the
action depend only on the zero-momentum part of the
perturbation. This also implies there is no contribution from
$S_{deriv}$ at linear order.  There is a potential divergence in the
action from the region at large $r$, where
\begin{equation}
A^2 = -\frac{2(z-1)}{z} (1+2\hat{a}_t), \quad F^2= -4z(z-1) (1 + 2\hat{a}_t + \frac{2r}{z} \partial_r \hat{a}_t + \frac{r}{z} \partial_r \hat{h}_{tt}).
\end{equation}
From the metric perturbation we have $\sqrt{-g} = r^{z+1}
[1+\frac{1}{2}(\hat{h}_{tt} + \hat{h}^i_{\ i})]$, $\sqrt{-h} = r^{z+2}
[1+\frac{1}{2}(\hat{h}_{tt} + \hat{h}^i_{\ i})]$, where
  $\hat{h}^i_{\ i} = \delta^{ij} \hat{h}_{ij}$,
\begin{equation}
R = -2z^2-4z-6 -r^2 \partial_r^2 \hat{h}_{tt} -  r^2 \partial_r^2
  \hat{h}^i_{\ i} - (2z+3) r \partial_r  \hat{h}_{tt} - (z+4)
  r \partial_r \hat{h}^i_{\ i},
\end{equation}
and
\begin{equation}
K = z+2 + \frac{r}{2} \partial_r (\hat{h}_{tt}+ \hat{h}^i_{\ i}).
\end{equation}
Putting all of this into the action \eqref{action} for a region $r
\leq r_0$ gives
\begin{eqnarray} \label{homact}
\frac{S}{\rm Vol} &=& bulk + \int^{r_0} dr r^{z+1} \left[-(z+2)(\hat{h}_{tt}+
\hat{h}^i_{\ i}) +2(z+2)(z-1)\hat{a}_t \right. \\ && \left. -(z+4) r \partial_r
(\hat{h}_{tt}+ \hat{h}^i_{\ i}) +2(z-1) r \partial_r \hat{a}_t -
r^2 \partial_r^2 (\hat{h}_{tt}+ \hat{h}^i_{\ i}) \right]  \nonumber \\
&&  + R^4 \left[\hat{h}_{tt}+ \hat{h}^i_{\ i} -2(z-1)\hat{a}_t +
r \partial_r (\hat{h}_{tt}+ \hat{h}^i_{\ i}) \right]_{r=r_0}, \nonumber
\end{eqnarray}
where we have performed the integral over $t,x,y$ and divided out the
overall volume in these directions. We write {\it ``bulk''} to
indicate that we are only keeping track of the contribution to the
action from the region at large $r$, where a linearised analysis is
appropriate. In the next subsection, we will determine the asymptotic
behaviour of these constant scalar perturbations, and show that the
potential divergences in the contributions we have written explicitly
in \eqref{homact}, coming from the region at large $r$, cancel to
leave a finite result.

We also want to verify that the variation of the action vanishes
on-shell for suitable boundary conditions on the variations. Our
approach will be to verify this by showing that the stress tensor
defined above is finite. The logic is that we can write the general
on-shell variation of the action as in \eqref{nrels}, with the
variation $\delta A^A$ restricted to a variation of $\delta A^0$ by
our choice of frame. If the action has finite variations under
variations which change the boundary data, the variation will then
clearly vanish for any variations that do not change the boundary data
(i.e., those which fall off fast enough at the boundary). We will show
below as we analyse the perturbations that they give finite
coefficients for variations of the boundary data, up to some
subtleties in the variation of $A^0$. These subtleties are addressed
in section \ref{vanish}, showing that the variation of the action
vanishes for suitable asymptotically Lifshitz boundary conditions.

Consider therefore the calculation of the non-relativistic stress
tensor complex defined in section \ref{stress} at linear order. Since
the $s_{\alpha\beta}$ and $s_\beta$ are already linear in terms of the
perturbation, our prescription for the stress tensor complex reduces
to
\begin{equation}
{\mathcal E} = -2 r^{-2z} s_{tt} + \alpha r^{-z} s_t, \quad {\mathcal
  E}_i = 2 r^{-2} s_{ti} - \alpha r^{z-2} s_i,
\end{equation}
\begin{equation}
{\mathcal P}_i = 2 r^{-2z} s_{ti}, \quad \Pi_{ij} = -2 r^{-2} s_{ij}.
\end{equation}
For the general perturbations, we should now include contributions
from $S_{deriv}$. We discuss this part of the calculation in appendix
\ref{deriv}. The upshot of the analysis there is that the
contributions from the derivative terms are suppressed relative to the
contribution from the non-derivative part of the action, and as a
result only make a finite contribution to the component $\mathcal E_y$
in the stress tensor complex, where they can be chosen to cancel
divergences in the contributions from the non-derivative part of the
action.

We can write the contribution from the remaining part of the action
for our ansatz in a relatively simple form in terms of the asymptotic
fields:
\begin{eqnarray}
{\mathcal E} &=&  -r^{z+2} \left[ r \partial_r
  \hat{h}^i_{\ i} + \alpha^2 ( z  \hat{a}_t +  r \partial_r
(\frac{1}{2} \hat{h}_{tt} + \hat{a}_t) - r^{-z} \partial_t \hat{a}_r)
\right] +  {\mathcal E}^{deriv} ,\\ \nonumber
{\mathcal E}_i &=& r^{z+2} \left[ r \partial_r v_{2i} +
  \frac{(z-2)}{z} r^{2(z-1)} r \partial_r v_{1i} - \frac{2(z-1)}{z}
  r^{z-2} \partial_i \hat{a}_r \right] + {\mathcal E}_i^{deriv},\\
\nonumber
{\mathcal P}_i &=& r^{z+2} [-r \partial_r v_{1i} + r^{-2(z-1)}
r \partial_r v_{2i} ]+ {\mathcal P}_i^{deriv}, \\ \nonumber
\Pi_{ij} &=& - r^{z+2} [ -r \partial_r \hat{h}_{tt}
\delta_{ij} + r \partial_r (\hat{h}_{ij} - \delta_{ij} \hat{h}^k_{\ k}) +
2(z-1) \hat{a}_t \delta_{ij} ] + \Pi_{ij}^{deriv}.
\end{eqnarray}
We will also want to evaluate
\begin{equation}
  s_0 = -r^{-z}  s_t = r^{z+2} \alpha [z \hat{a}_t + r \partial_r
  (\frac{1}{2} \hat{h}_{tt} + \hat{a}_t) - r^{-z} \partial_t \hat{a}_r
  ] + s_0^{deriv}.
\end{equation}
For completeness, we also note that
\begin{equation}
s_i = -r^{z+2} \alpha [r^z r \partial_r v_{1i} - \partial_i
\hat{a}_r].
\end{equation}

In our linearised analysis, terms in the conservation equations
involving the variation of the matter fields like the one appearing on
the right-hand side of \eqref{relcons} will not appear, as both $s_A$
and the derivative $\partial_\beta A^A$ are of linear order in the
perturbation. We therefore expect our stress tensor complex to obey
the conservation equations \eqref{cons}, and we will indeed find that
the bulk equations of motion imply this conservation.

Finally, a note on the applicability of this linearised analysis. We
can see from the form of the stress tensor that perturbations where
the normalised fields fall off like $r^{-(z+2)}$ will be associated
with finite contributions to some element of the stress tensor
complex. Thus, if we have linear perturbations where the normalised
fields fall off like $r^{-\frac{1}{2}(z+2)}$, then quadratic terms in
these perturbations could make finite contributions to the stress
tensor complex, and the linearised analysis we are performing would
not be justified by the smallness of the fields in the asymptotic
region; even to understand the asymptotic behaviour of a generic
asymptotically Lifshitz solution with such falloffs could require a
non-linear analysis.

\subsection{Constant perturbations}

Because the background is translation-invariant in $t,x,y$, we can
decompose the perturbations into plane wave modes, and modes of
different frequencies will not mix. We consider first the zero
momentum part; perturbations which are constant in the boundary
directions. These constant perturbations can be decomposed
into scalar, vector and tensor parts:
\begin{equation}
  h_{tt} = - r^{2z} f(r), \quad h_{ti} = -r^{2 z} v_{1i}(r) + r^2
  v_{2i}(r), \quad h_{ij} = r^2 k(r)  \delta_{ij} + r^2 k_{ij}(r),
\end{equation}
where
\begin{equation}
k_{ij}(r) = \left[ \begin{array}{cc} t_d(r) & t_o(r) \\ t_o(r) & -t_d(r)
  \end{array} \right],
\end{equation}
and
\begin{equation}
a_t = \alpha r^z (j(r) + \frac{1}{2} f(r)), \quad a_i = \alpha r^z
v_{1i}(r).
\end{equation}
Note that a constant $a_r$ component is forced to vanish by the
equations of motion. As a consequence of the rotation invariance in
the $x,y$ plane in the background, the scalar, vector and tensor
sectors do not mix.

We consider first the constant scalar perturbations. This will include
as a special case the linearised version of the black hole solutions
of \cite{dt,mann}. While this paper was in preparation, the
perturbations in this scalar sector were analysed in \cite{peet},
which also considers a background where the flat spatial slices are
replaced by a sphere. Our results agree with this previous work,
although direct comparison is not straightforward as we work in a
different gauge.

The equations of motion for constant scalar modes reduce to
\begin{eqnarray}
2r^2 j'' &=& (z+1) r f' -4 (z+1) r j' - (z+4) (2z-2) j, \\
\frac{1}{r^2} (z+1) (r^4 f')' &=& (z-1)(4z+2) rj' + (z-1)(4z^2+6z+8) j, \\
2(z+1) rk' &=& -(z+1) r f' - 2(z-1) rj' -(z-1)(2z-4) j.
\end{eqnarray}
The fact that these do not involve $f, k$ undifferentiated reflects
the freedom to shift coordinates by rescaling $t, x, y$.

For $z=2$, the solution is
\begin{eqnarray}
j(r) &=& -\frac{c_1 + c_2 \ln r}{r^4}+c_3 , \\
f(r) &=& \frac{4 c_1 - 5 c_2 + 4 c_2 \ln r}{12 r^4} + 4c_3 \ln r +
c_4, \\
k(r) &=& \frac{4 c_1 + 5 c_2 + 4 c_2 \ln r}{24 r^4} - 2c_3 \ln r +
c_5.
\end{eqnarray}
We can set $c_4= c_5 = 0$ by redefining the coordinates $t,x,y$. We
should also set $c_3=0$ to satisfy the asymptotically Lifshitz
boundary condition; that is, to ensure that the solution is small at
large $r$, consistent with our assumption.

For $z \neq 2$, the solution is
\begin{eqnarray}
j(r) &=& -\frac{(z+1)c_1}{(z-1)r^{z+2}} -
\frac{(z+1)c_2}{(z-1)r^{\frac{1}{2}(z+2+\beta_z)}} +
\frac{(z+1)c_3}{{(z-1)r^{\frac{1}{2}(z+2-\beta_z)}}}, \\
f(r) &=& 4\frac{1}{(z+2)} \frac{c_1}{r^{z+2}} + 2
\frac{(5z-2-\beta_z)}{(z+2+\beta_z)}
\frac{c_2}{r^{\frac{1}{2}(z+2+\beta_z)}}  \\ && -2
\frac{(5z-2+\beta_z)}{(z+2-\beta_z)}
\frac{c_3}{{r^{\frac{1}{2}(z+2-\beta_z)}}} + c_4, \nonumber \\
k(r) &=& 2 \frac{1}{(z+2)} \frac{c_1}{r^{z+2}} - 2
\frac{(3z-4-\beta_z)}{(z+2+\beta_z)}
\frac{c_2}{r^{\frac{1}{2}(z+2+\beta_z)}} \\ && +2
\frac{(3z-4+\beta_z)}{(z+2-\beta_z)}
\frac{c_3}{{r^{\frac{1}{2}(z+2-\beta_z)}}} +c_5, \nonumber
\end{eqnarray}
where $\beta_z^2 = 9z^2-20z+20 = (z+2)^2 +8(z-1)(z-2)$.

Let us use these solutions to be more precise about the asymptotically
Lifshitz boundary conditions. We see that there are constant modes in
$f$ and $k$, which can be interpreted as changes in the boundary data
for the metric. For $j$, by contrast, there is no constant mode for
general $z$. The slowest falloff in $j$ is given by the mode
parametrized by $c_3$, which falls off as
$r^{-\frac{1}{2}(z+2-\beta_z)}$ (it is constant in the special case
$z=2$). Thus, for $1 \leq z <2$, we want to interpret the mode
parametrised by $c_3$ as the boundary data for the vector field. To
fix this boundary data, we need to require that
$r^{\frac{1}{2}(z+2-\beta_z)}( A^M - \alpha \delta^M_{\ 0})$ vanishes
as $r \to \infty$.\footnote{Note that this implies, surprisingly, that
  the boundary data are subleading compared to the background value
  for $A^M$.  For general $z$, the allowed changes in the boundary
  data for the massive vector do not change the $\alpha \delta^M_{\
    0}$ term, but add a term falling off like
  $r^{-\frac{1}{2}(z+2-\beta_z)}$ to it. Apart from the subtraction of
  the $\alpha \delta^M_{\ 0}$ term, this is like the boundary
  condition for a massive vector in the relativistic case.  } For $z
\geq 2$, this mode produces terms in $f$ and $k$ which grow at large
$r$, and hence violate the boundary conditions for those fields. It is
therefore not clear whether we can think of this as boundary data for
the vector field in this case. For $z \geq 2$, we will simply impose
the boundary condition that $A^M - \alpha \delta^M_{\ 0}$ vanishes as
$r \to \infty$. We therefore adopt as our definition of asymptotically
Lifshitz boundary conditions that $\hat{h}_{tt}$, $v_{1i}$, $v_{2i}$,
$\hat{h}_{ij}$ and $\hat a_t$ vanish as $r \to \infty$, and that for
$1 \leq z < 2$, $r^{\frac{1}{2}(z+2-\beta_z)} \hat a_t \to 0$ as $r
\to \infty$.

We thus have a two-parameter family of solutions in this constant
scalar sector, parametrized by $c_1, c_2$. In ~\cite{peet}, the
energy for these solutions was evaluated by background subtraction,
and they found that for $z \leq 2$, they needed to set $c_2 = 0$ as
well to have a finite energy. We will see below that with our
definition of the boundary energy density, we get finite results for
any $z$ without further restricting the solutions.\footnote{ In fact,
  for constant modes, we have a finite energy even if we allow $c_3
  \neq 0$.} The divergences found in~\cite{peet} are due to using an
action which does not include the surface terms necessary to ensure
the action is finite on-shell. In the cases $z \leq 2$, the
asymptotically Lifshitz solution approaches the background too slowly
at large $r$ for these surface terms to cancel out in the background
subtraction calculation. A similar failure of background subtraction
occurs for the Schr\"odinger case~\cite{heat,Azeyanagi:2009pr}.

We first want to use these scalar modes to evaluate the on-shell value
of the action \eqref{homact}.  For $z=2$, we find
\begin{equation}
\frac{S}{\rm Vol} = bulk + \frac{2}{3} c_2,
\end{equation}
and for $z \neq 2$, we find
\begin{equation}
\frac{S}{\rm Vol} = bulk + \frac{2(z+1)(z-2)}{(z+2)} c_1.
\end{equation}
Thus, we see that the potential divergences from the region at large
$r$ cancel, to leave a finite answer for this part of the action. We
have not explicitly considered the contribution to the action from the
interior of the spacetime, so there could still be a divergence there,
but this is unlikely. In appendix \ref{Euclidean}, we consider the action
for a set of static Euclidean black hole solutions, and see explicitly
that it is finite.

We next consider the contribution to the stress tensor from these
modes, which gives
\begin{equation}
{\mathcal E} = -r^{z+2}[ 2r \partial_r k + \alpha^2 (z j + r
  \partial_r j +  \frac{1}{2} r \partial_r f)] = \left\{
  \begin{array}{cc}
  \frac{4(z-2)}{z} c_1 & \mbox{ for }z\neq 2 \\  \frac{4
  c_2}{3}   & \mbox{ for }z = 2. \end{array} \right.
\end{equation}
Note that the separate contributions are divergent (logarithmically
for $z=2$), but the combination is finite. We have
\begin{equation}
\Pi_{ij} = -2r^{-2} s_{ij} = -2r^{z+2}[(z-1)j - \frac{r}{2} \partial_r f -
  \frac{r}{2} \partial_r k ] \delta_{ij}= \left\{ \begin{array}{cc}
  2(z-2) c_1 \delta_{ij} & \mbox{ for }z\neq 2 \\  \frac{4
  c_2}{3}  \delta_{ij} & \mbox{ for }z = 2. \end{array} \right.
\end{equation}
For the black hole solutions \cite{dt,mann}, only these scalar modes
are turned on, so this gives the thermal stress tensor dual to the
black hole solution. The bulk black hole can be used to relate the
energy density, which is an arbitrary constant of integration in our
asymptotic analysis, to the temperature. Note that this thermal stress
tensor satisfies the equation of state $z{\mathcal E} = \delta^{ij}
\Pi_{ij}$ required by the anisotropic scaling symmetry.

For the vector modes, the equations are
\begin{eqnarray}
r^2 v_{1i}'' + (2z+1) r v_{1i}' + z r^{-2(z-1)} r v_{2i}' &=& 0, \\
r^2 v_{2i}'' + 5 r v_{2i}' + (z-2) r^{2(z-1)} r v_{1i}' &=& 0.
\end{eqnarray}
For $z\neq4$ the solutions are
\begin{eqnarray}
v_{1i}(r) &=& c_{1i} + \frac{c_{2i}}{r^{z+2}} + \frac{c_{3i}}{r^{3z}},
\\
v_{2i}(r) &=& \frac{(z^2-4)}{z(z-4)} c_{2i} r^{z-4} +
\frac{3z}{(z+2)}  \frac{c_{3i}}{r^{z+2}} +
c_{4i}, \nonumber
\end{eqnarray}
and for $z=4$ we have
\begin{eqnarray}
v_{1i}(r) &=& c_{1i} + \frac{c_{2i}}{r^{6}} + \frac{c_{3i}}{r^{12}},
\\
v_{2i}(r) &=& 3\ln(r)c_{2i}+
2\frac{c_{3i}}{r^{6}} +
c_{4i}. \nonumber
\end{eqnarray}
These give contributions to the stress tensor complex which are
\begin{equation}
{\mathcal E}_i = r^{z+2} \left[ r \partial_r v_{2i} +
  \frac{(z-2)}{z} r^{2(z-1)} r \partial_r v_{1i} \right] =-6 (z-1) c_{3i},
\end{equation}
and
\begin{equation}
{\mathcal P}_i = r^{z+2} [-r \partial_r v_{1i} + r^{-2(z-1)}
r \partial_r v_{2i} ] = \frac{2(z-1)(z+2)}{z} c_{2i}.
\end{equation}

In the vector solutions, $c_{1i}$ is a pure gauge mode corresponding
to shifting $t \to t+ c_{1i} x^i$, and $c_{4i}$ is a pure gauge mode
corresponding to shifting $x^i \to x^i + c_{4i} t$. The contribution
of $c_{2i}$ to $v_{2i}$ falls off more slowly than $r^{-(z+2)/2}$ for
$z > 2$, so we would expect a linearised analysis to be insufficient
to correctly extract the boundary stress tensor for a generic
asymptotically Lifshitz geometry for $z \geq 2$.  For $z \geq 4$, we
need to set the coefficient $c_{2i}$ to zero to satisfy the boundary
conditions on $v_{2i}$. This is a further restriction on the space of
allowed solutions, which makes the space of allowed solutions (at
least in the linearised approximation) two dimensions smaller. This
restriction sets ${\mathcal P}_i =0$ for all asymptotically
Schr\"odinger solutions with $z \geq 4$. It would be very interesting
to understand this restriction from the dual field theory point of
view. It would also be interesting to see if one can construct
solutions with a non-zero boost that are physically acceptable and at
the same time have $c_{2i}=0$.

For the tensor modes, the equations are
\begin{eqnarray}
r^2 t_d'' + (z+3) r t_d' &=& 0, \\
r^2 t_o'' + (z+3) r t_o' &=& 0,
\end{eqnarray}
and the solutions are
\begin{equation}
t_d(r) = t_{d1} + \frac{t_{d2}}{r^{z+2}}, \quad t_o(r) = t_{o1} +
\frac{t_{o2}}{r^{z+2}}.
\end{equation}
The constant terms are pure gauge, corresponding to relative
scaling and rotation of the $x,y$ coordinates respectively.
The tensor modes source
\begin{equation}
\Pi_{ij} = - r^{z+2} r \partial_r k_{ij},
\end{equation}
which gives
\begin{equation}
\Pi_{xy} =(z+2) t_{o2}, \quad \Pi_{xx} = -\Pi_{yy} =
(z+2) t_{d2}.
\end{equation}

Since all of the constant perturbation modes give constant components
for the stress tensor, the conservation equations are trivially
satisfied.

In summary, for the constant perturbations, we have an eight-parameter
family of solutions of the linearised equations of motion satisfying
our asymptotic boundary conditions. Seven of these parameters
correspond to the independent components of the stress tensor complex;
there is an additional linearised solution in the scalar sector which
does not contribute to the stress tensor at this order. For $1\le z < 2$,
we have tightened our boundary conditions by setting $c_3 = 0$ even
though this mode does not grow asymptotically. For $z \geq 4$, we must
set ${\mathcal P}_i = 0$ to satisfy our boundary conditions, and we
have a six-parameter family of solutions in the bulk.

\subsection{General perturbations}
\label{gpert}

Taking a particular plane wave mode, we can again decompose the
perturbation into scalar and vector parts (for non-zero momentum,
there is no transverse tracefree tensor in two dimensions). We
simplify the analysis by using the rotation invariance to take the
momentum to lie only along the $x$ direction. Then we can
write\footnote{To avoid cluttering the notation, we will not
  introduce subscripts $\omega,k$ on the functions in this ansatz
  to denote the mode we are considering. We hope this will not lead to
confusion.}
\begin{equation}
h_{tt} = - r^{2z} f(r) e^{i\omega t+ i k x}, \quad h_{tx} =
  k [-r^{2z} s_1(r) + r^2 s_2(r)] e^{i\omega t+ i
  k x} ,
\end{equation}
\begin{equation}
 h_{ty} = [-r^{2z} v_1(r) + r^2 v_2(r)] e^{i\omega t+ i
  k x},
\end{equation}
\begin{equation}
h_{xx} = r^2 \left( k_L(r) +  k^2
  k_T(r) \right)e^{i\omega
  t+ i k x} , \quad h_{yy} = r^2 \left( k_L(r) -  k^2
  k_T(r) \right)e^{i\omega
  t+ i k x},
\end{equation}
\begin{equation}
h_{xy} = r^2 k v_3(r) e^{i\omega
  t+ i k x},
\end{equation}
and
\begin{equation}
a = \alpha r^z e^{i\omega t+ i k x} [(j(r) + \frac{1}{2} f(r)) dt +
k s_1(r) dx  + v_1(r) dy + i\frac{p(r)}{r^{z+1}} dr].
\end{equation}
The functions $v_1, v_2, v_3$ represent divergence-free vector
excitations, while the other functions are scalars or scalar-derived
vectors with respect to the rotational symmetry in the $x,y$
plane. The scalar modes and vector modes decouple, so we can analyse
them separately.

\subsubsection{Scalar modes}

For the scalar part, one can bring the equations of motion to a nicer
form by rescaling $s_1 \to \omega s_1$, $s_2 \to \omega s_2$. The
function $p(r)$ appearing in $a_r$ is determined algebraically,
\begin{equation}
p(r) = \frac{\omega}{4(z-1)r^z} \left[ -2r k_L' +2(z-1) k_L + k^2 (r
  s_2' - r^{2z-1} s_1' - 2(z-1) s_2) \right]
\end{equation}
and using this to eliminate $p(r)$, the remaining equations of motion
for the modes in the scalar sector are
\begin{eqnarray} \label{fj1}
-2(z-1) r j' + (z+1) r f' - 6z(z-1) j &=&  -\frac{k^2}{r^2} (-2(z+1) r^3
k_T' + k_L + f - k^2 k_T) \\ \nonumber && - \frac{\omega^2}{r^{2z}} (r k_L' + 2 (z+1)
r^3 ( s_2' - r^{2(z-1)} s_1') - (z+1) k_L ) \\ \nonumber &&- \frac{\omega^2
  k^2}{2r^{2z}} ( -r s_2' + r^{2z-1} s_1' +2(z+1) s_2 -4 r^{2(z-1)} s_1
),
\end{eqnarray}
\begin{eqnarray} \label{fj2}
r^2 f'' - (2z-3) rf' + 8(z-1)^2 j &=& \frac{k^2}{r^2} (-2(2z+1) r^3 k_T'
+ k_L + 2f - k^2 k_T) \\ \nonumber &&+ \frac{\omega^2}{r^{2z}} (2(2z+1) r^3 (s_2'
- r^{2(z-1)} s_1') -4 k_L) \\ \nonumber &&+ \frac{4k^2 \omega^2}{r^{2z}} (s_2
- r^{2(z-1)} s_1),
\end{eqnarray}
\begin{equation} \label{kl}
r k_L' + r f' - 2(z-1) j = k^2 r k_T' + \omega^2 (rs_1' - r^{-2(z-1)}
r s_2'),
\end{equation}
\begin{equation} \label{kt}
r^4 k_T'' + (z+3) r^3 k_T' - \frac{1}{2} f = -\omega^2 \left[ s_1 +
  r^{-2(z-1)} (k_T - s_2) \right],
\end{equation}
\begin{eqnarray}\label{vec2}
-2r ^4 s_2'' + 2 r^{2z+2} s_1'' +2 (z-5) r^3 s_2' + 2(z+3) r^{2z+1}
s_1'\\  -2 r f' + 4(z-1) j -2 (z-2) k_L \nonumber \\ = k^2(r s_2'
-r^{2z-1} s_1' -2 r k_T' -2(z-1) s_2 + 2 k_T) + 2\omega^2 (r^{-2z+3}
s_2' - r s_1'),   \nonumber
\end{eqnarray}
\begin{eqnarray}\label{vec1}
k^2 (-2r f' -4 r^3 k_T' -2(z-2) r^{2z+1} s_1' -2z r^3 s_2' +2f +
4(z-1) j-2(z-2) k_L) &&\\ \nonumber + \frac{\omega^2}{r^{2(z-1)}} ( 4r^3 s_2' -4
  r^{2z+1} s_1' -4k_L)  && \\  \nonumber + \frac{k^2 \omega^2}{r^{2z}}
  (2r^{2z+1} s_1'   - 2 r^3 s_2' +4 r^2 s_2 - 4 r^{2z} s_1) && \\
  \nonumber + k^4
  (r^{2z-1} s_1' -r s_2'  + 2r k_T' + 2(z-1) s_2 -2 k_T) &=& 0.
\end{eqnarray}
For the scalar modes, the non-zero contributions to the boundary
stress tensor complex are
\begin{eqnarray} \label{glstress}
{\mathcal E} &=&  -r^{z+2} \left[ 2 r \partial_r k_L + \alpha^2 ( z  j
  +  r \partial_r (\frac{1}{2} f + j ) + \omega r^{-z} p) \right]
e^{i \omega t + i kx}+ {\mathcal E}^{deriv},\\ \nonumber
{\mathcal E}_x &=& r^{z+2} \left[ k\omega r \partial_r s_{2} +
  k\omega \frac{(z-2)}{z} r^{2(z-1)} r \partial_r s_{1} + \frac{2(z-1)}{z}
  k r^{z-2} p \right]e^{i \omega t + i kx} + {\mathcal E}_x^{deriv},\\ \nonumber
{\mathcal P}_x &=& r^{z+2} [-k\omega r \partial_r s_{1} + k\omega r^{-2(z-1)}
r \partial_r s_{2} ]e^{i \omega t + i kx}+ {\mathcal P}_x^{deriv}, \\
\nonumber
\Pi_{xx} &=& - r^{z+2} [ -r \partial_r f +  r \partial_r
(-k_L + k^2 k_T) +
2(z-1) j ]e^{i \omega t + i kx} + \Pi_{xx}^{deriv}, \\
\nonumber
\Pi_{yy} &=& - r^{z+2} [ -r \partial_r f + r \partial_r (-k_L - k^2 k_T) +
2(z-1) j ]e^{i \omega t + i kx} + \Pi_{yy}^{deriv}.
\end{eqnarray}
We see in appendix \ref{deriv} that the derivative terms make a
vanishing contribution for the scalar modes.

The full stress tensor is conserved by virtue of the bulk equations of
motion: for the terms in \eqref{glstress}, the $xr$ component of
Einstein's equation gives the conservation equation $\omega {\mathcal
  P}_x + k \Pi_{xx} = 0$, and a combination of the $tr$ component of
Einstein's equation and the $r$ component of the massive vector
equation gives the conservation equation $\omega {\mathcal E} + k
{\mathcal E}_x = 0$.

We solve (\ref{fj1}-\ref{vec1}) by writing each of the functions in a power
series in $\omega$, $k$. If we denote the functions collectively by
$F$, we have $F = \sum_{l,m} k^{2l} \omega^{2m} F^{(l,m)}$. The
equations for the $(0,0)$ part of the functions are obtained by taking
the $k^0 \omega^0$ part of (\ref{fj1}-\ref{kt}) and the $k^2$ and
$\omega^2$ parts of \eqref{vec1} (which imply the $k^0 \omega^0$ part
of \eqref{vec2}). The equations are then
\begin{eqnarray}
-2(z-1) r j^{(0,0)'} + (z+1) r f^{(0,0)'} - 6z(z-1) j^{(0,0)} &=& 0,\\
 r^2 f^{(0,0)''} - (2z-3) rf^{(0,0)'} + 8(z-1)^2 j^{(0,0)} &=& 0, \\
r k_L^{(0,0)'} + r f^{(0,0)'} - 2(z-1) j^{(0,0)} &=& 0, \\
 r^4 k_T^{(0,0)''} + (z+3) r^3 k_T^{(0,0)'} - \frac{1}{2} f^{(0,0)} &=& 0, \\
2(z-1) r^3 s_2^{(0,0)'} + 2 r^3 k_T^{(0,0)'} + r f^{(0,0)'} -f^{(0,0)} -2(z-1) j^{(0,0)} &=& 0, \\
2(z-1) r^{2(z-1)} r^3 s_1^{(0,0)'} + 2 r^3 k_T^{(0,0)'} && \nonumber \\ + r f^{(0,0)'} - f^{(0,0)} + 2(z-1) k_L^{(0,0)} -2
(z-1) j^{(0,0)} &=& 0.
\end{eqnarray}
The solution of the first two equations for $f$, $j$ is
\begin{equation}
j^{(0,0)} =
\frac{(z+1)d_2}{(z-1)r^{\frac{1}{2}(z+2+\beta_z)}} +
\frac{(z+1)d_3}{(z-1)r^{\frac{1}{2}(z+2-\beta_z)}},
\end{equation}
\begin{equation}
\quad f^{(0,0)} = -\frac{2(5z-2-\beta_z)}{(z+2+\beta_z)}
\frac{d_2}{r^{\frac{1}{2}(z+2+\beta_z)}} -
\frac{2(5z-2+\beta_z)}{(z+2-\beta_z)}
\frac{d_3}{r^{\frac{1}{2}(z+2-\beta_z)}} + d_4
\end{equation}
for $z \neq 2$, and
\begin{equation}
j^{(0,0)} = \frac{3d_2}{r^4} + d_3, \quad f^{(0,0)} =
-\frac{d_2}{r^4} + 4 d_3 \ln r + d_4
\end{equation}
for $z=2$.  In the other functions, in addition to the terms sourced
by these modes, there is an arbitrary constant term in $k_L$, and a
solution for $k_T^{(0,0)} = d_5 + d_1 r^{-z-2}$. These source terms in
$s_1$ and $s_2$, which also have arbitrary constant terms. Set all the
constant terms to zero to satisfy the asymptotic boundary conditions,
and also set $d_3=0$ as in the discussion of the constant modes.

We are then left with two solutions of the coupled system: the first
is\footnote{We are introducing new constants $c_i$ here to parametrize
  the independent solutions which satisfy the asymptotic boundary
  conditions. The $s_1^{coeff}$, $s_2^{coeff}$ are unimportant but
  complicated numerical factors, so we do not write them explicitly.}
\begin{equation}
j^{(0,0)} = \frac{(z+1)c_1}{(z-1)r^{\frac{1}{2}(z+2+\beta_z)}}, \quad f^{(0,0)} = -
\frac{2(5z-2-\beta_z)}{(z+2+\beta_z)}
\frac{c_1}{r^{\frac{1}{2}(z+2+\beta_z)}},
\end{equation}
\begin{equation}
k_L^{(0,0)} =\frac{2(3z-4-\beta_z)}{(z+2+\beta_z)}
\frac{c_1}{r^{\frac{1}{2}(z+2+\beta_z)}},
\end{equation}
\begin{equation}
k_T^{(0,0)} = -
\frac{(z+1)(5z-2-\beta_z)}{2(z+2+\beta_z)(z^2-3z+4+\beta_z)}
\frac{c_1}{r^{\frac{1}{2}(z+6+\beta_z)}},
\end{equation}
\begin{equation}
s_1^{(0,0)} = s_1^{coeff} \frac{c_1}{r^{2z} r^{\frac{1}{2}(z+2+\beta_z)}}, \quad
s_2^{(0,0)} = s_2^{coeff} \frac{c_1}{r^2 r^{\frac{1}{2}(z+2+\beta_z)}}
\end{equation}
for $z \neq 2$, and
\begin{equation}
j^{(0,0)} = \frac{3c_1}{r^4}, \quad f^{(0,0)} = -\frac{c_1}{ r^4}, \quad
k_L^{(0,0)} = -\frac{c_1}{2 r^4}, \quad k_T^{(0,0)} = -\frac{c_1}{24 r^6},
\end{equation}
\begin{equation}
s_1^{(0,0)} = -\frac{3c_1}{32 r^8}, \quad s_2^{(0,0)} = -\frac{c_1}{24 r^6}
\end{equation}
for $z=2$.

This first solution will not give a contribution to the stress
tensor. For $z>2$, its contribution is a negative power of $r$, so it
vanishes in any case. For $z \leq 2$, the contribution of this
leading-order part is a non-negative power of $r$, but an explicit
calculation shows that the coefficient vanishes, as for the constant
perturbations. As $\beta_{z} -(z+2) > -1/2$, the first subleading piece,
which is suppressed by $k^2/r^2$ relative to the leading pieces, will
always give a negative power of $r$, so we do not need to compute
it. Thus, the mode parametrized by $c_1$ makes zero
contribution to the stress tensor complex.

The other solution of the leading-order equations satisfying our
boundary conditions is
\begin{equation}
k_T^{(0,0)} = -\frac{c_2}{r^{z+2}}, \quad s_1^{(0,0)} = \frac{(z+2) c_2}{3 z(z-1)
  r^{2(z-1)} r^{z+2}}, \quad s_2^{(0,0)} = \frac{c_2}{(z-1) r^{z+2}}.
\end{equation}
This will make a finite contribution to the stress tensor complex. To
calculate it fully, we need to first calculate some of the
higher-order terms in our expansion.

Next we consider the solution for the functions $F^{(1,0)}$. The
equations determining these functions will be the $k^2$ components of
(\ref{fj1}-\ref{vec2}) and the $k^4$ component of \eqref{vec1}. These
equations are
\begin{equation}
-2(z-1) r j^{(1,0)'} + (z+1) r f^{(1,0)'} - 6z(z-1) j^{(1,0)} =
 -\frac{1}{r^2} (-2(z+1) r^3
k_T^{(0,0)'} + k_L^{(0,0)} + f^{(0,0)}),
\end{equation}
\begin{equation}
 r^2 f^{(1,0)''} - (2z-3) rf^{(1,0)'} + 8(z-1)^2 j^{(1,0)} =
 \frac{1}{r^2}(-2(2z+1) r^3 k_T^{(0,0)'}
+ k_L^{(0,0)} + 2f^{(0,0)}),
\end{equation}
\begin{equation}
r k_L^{(1,0)'} + r f^{(1,0)'} - 2(z-1) j^{(1,0)} = r k_T^{(0,0)'},
\end{equation}
\begin{equation}
 r^4 k_T^{(1,0)''} + (z+3) r^3 k_T^{(1,0)'} - \frac{1}{2} f^{(1,0)} = 0,
\end{equation}
\begin{eqnarray}
&-2r ^4 s_2^{(1,0)''} + 2 r^{2z+2} s_1^{(1,0)''} +2 (z-5) r^3 s_2^{(1,0)'} + 2(z+3) r^{2z+1}
s_1^{(1,0)'} \\ \nonumber &-2 r f^{(1,0)'} + 4(z-1) j^{(1,0)} -2 (z-2)
k_L^{(1,0)} \\ \nonumber &=(r s_2^{(0,0)'}
-r^{2z-1} s_1^{(0,0)'} -2 r k_T^{(0,0)'} -2(z-1) s_2^{(0,0)} + 2
k_T^{(0,0)})
\end{eqnarray}
\begin{eqnarray}
&-2r f^{(1,0)'} -4 r^3 k_T^{(1,0)'} -2(z-2) r^{2z+1} s_1^{(1,0)'} -2z
  r^3 s_2^{(1,0)'} \\ \nonumber &+2f^{(1,0)} +
4(z-1) j^{(1,0)}-2(z-2) k_L^{(1,0)} = \\ \nonumber &-
  (r^{2z-1} s_1^{(0,0)'} -r s_2^{(0,0)'}  + 2r k_T^{(0,0)'} + 2(z-1) s_2^{(0,0)} -2 k_T^{(0,0)}).
\end{eqnarray}

This system will have a homogeneous solution of the same form as the
solution of the $F^{(0,0)}$ equations; we can absorb that into the
$F^{(0,0)}$ solution by a redefinition of $c_1,c_2$. We will absorb all
homogeneous solutions of the same form at higher orders in the same
way, promoting these constants to arbitrary functions of
$k,\omega$. Because the equations for $s_1$ and $s_2$ are different,
there is an additional homogeneous solution which did not appear in
the  $F^{(0,0)}$ solutions. This is
\begin{equation}
s_1^{(1,0)} = \frac{c_3}{r^{z+2}}, \quad s_2^{(1,0)} =
\frac{(z^2-4)}{z(z-4)} c_3 r^{z-4}.
\end{equation}
As in the constant perturbations, we must set $c_3 = 0$ for $z \geq 4$
to satisfy the asymptotic boundary condition that $s_2 \to 0$ as $r
\to \infty$.

In addition to the homogeneous solutions, we will have particular
integrals for the sources from the $F^{(0,0)}$ solutions. As we have
said above, there will be non-trivial particular integrals for the
solution parametrized by $c_1$, but they do not contribute to the stress
tensor, so we will not calculate them explicitly.
For the solution parametrized by $c_2$, a particular integral for $z
\neq 2$ is
\begin{equation} \label{d10}
f^{(1,0)} = \frac{2c_2}{(z-2) r^{z+2}}, \quad j^{(1,0)} =
-\frac{(z+2)(z+1)c_2}{2(z-2) (z-1) r^{z+2}}, \quad k_L^{(1,0)} =
\frac{c_2}{(z-2) r^{z+2}},
\end{equation}
\begin{equation}
k_T^{(1,0)} = \frac{c_2 }{2(z+4)(z-2)r^{z+4}},
\end{equation}
\begin{equation}
s_1^{(1,0)} =
\frac{3 c_2 }{2(z-1)(z-2)(3z+2) r^{2(z-1)} r^{z+4}}, \quad s_2^{(1,0)}
= -\frac{(z-4) c_2 }{2z (z-1)(z-2)(z+4) r^{z+4}}.
\end{equation}
For $z=2$, a particular integral is
\begin{equation}\label{d102}
j^{(1,0)} = -\frac{9 c_2 \ln r}{r^4}, \quad f^{(1,0)} =  \frac{3 c_2 \ln
  r}{r^4} +\frac{c_2}{4 r^4}, \quad k_L^{(1,0)} =  \frac{3 c_2 \ln r}{2
  r^4} - \frac{c_2}{8 r^4},
\end{equation}
\begin{equation}
k_T^{(1,0)} =  \frac{c_2 \ln r}{8 r^6} + \frac{3 c_2}{32 r^6}, \quad
s_1^{(1,0)} =  \frac{9 c_2 \ln r}{32 r^8} - \frac{93 c_2}{256 r^8}, \quad
s_2^{(1,0)} =  \frac{c_2 \ln r}{8 r^6} - \frac{13 c_2}{32 r^6}.
\end{equation}

Only the terms in \eqref{d10} or \eqref{d102} contribute to the stress
tensor. This solution will lead to further contributions in the higher
$F^{(l,m)}$, but they are suppressed by further powers of $r$, so they
do not contribute to the stress tensor complex. We can therefore
evaluate the contribution for this mode,
\begin{equation}
{\mathcal E} = \frac{2 (z+2)}{z} c_2 k^2e^{i \omega t + i kx}, \quad
{\mathcal E}_x = - \frac{2 (z+2)}{z} c_2 k \omega e^{i \omega t + i kx},
\end{equation}
\begin{equation}
{\mathcal P}_x= 0, \quad \Pi_{xx} = 0, \quad \Pi_{yy} = 2(z+2) c_2
k^2e^{i \omega t + i kx}.
\end{equation}
The conservation equation $\omega {\mathcal E} + k {\mathcal E}_x =0$
and the trace condition $z {\mathcal E} = \delta^{ij} \Pi_{ij}$
are satisfied as required.

We can carry on and calculate the equations of motion for the
$F^{(0,1)}$ functions. The relevant equations are the $\omega^2$ parts
of (\ref{fj1}-\ref{kt}) and the $k^2 \omega^2$ and $\omega^4$ parts of
\eqref{vec1}. As a result, the homogeneous solutions will be exactly
the same as for the $F^{(0,0)}$, and we are only interested in the
particular integrals which can contribute to the stress tensor. The
only relevant terms are the ones proportional to $c_3$. The only
source term from the $F^{(1,0)}$ functions in the equations for the
$F^{(0,1)}$ functions is in the equation obtained from the $k^2 \omega^2$ part of
\eqref{vec1},
\begin{eqnarray} \label{kt01}
&-2r f^{(0,1)'} -4 r^3 k_T^{(0,1)'} -2(z-2) r^{2z+1} s_1^{(0,1)'} -2z
  r^3 s_2^{(0,1)'} \\ \nonumber &+2f^{(0,1)} +
4(z-1) j^{(0,1)}-2(z-2) k_L^{(0,1)} \\ \nonumber &= -
\frac{1}{r^{2(z-1)}} ( 4r^3 s_2^{(1,0)'} -4
  r^{2z+1} s_1^{(1,0)'} -4k_L^{(1,0)}).
\end{eqnarray}
A particular integral which satisfies the full set of equations for
the $F^{(0,1)}$ functions is
\begin{equation}
k_T^{(0,1)} = - \frac{2(z-1) c_3}{z r^{z+2}}.
\end{equation}

At higher orders, there will be no new homogeneous solutions. The
homogeneous part of the equations for $F^{(l,m)}$ is the same as
$F^{(1,0)}$ for $l \neq 0$, and is the same as $F^{(0,0)}$ for
$l=0$.\footnote{The equations of motion for $F^{(0,m)}$ are in general
  the $\omega^{2m}$ part of (\ref{fj1}-\ref{kt}) and the $k^2
  \omega^{2m}$ and the $\omega^{2m+2}$ parts of \eqref{vec1}. One can
  check that in general the $\omega^{2m+2}$ part of \eqref{vec1}
  together with the $\omega^{2m}$ part of \eqref{kl} imply the
  $\omega^{2m}$ part of \eqref{vec2}.} Thus, we can absorb the
homogeneous solution into a redefinition of $c_1, c_2, c_3$. As for
the particular integrals, we have obtained all the terms involving
$c_1, c_2$ which can affect the stress tensor; higher terms are
suppressed. For the solutions involving $c_3$, there is no source term
in (\ref{fj1}-\ref{kt}) for the functions $f^{(2,0)}, j^{(2,0)},
k_L^{(2,0)}, k_T^{(2,0)}$. The solution is therefore simply
\begin{equation}
s_1^{(2,0)} = -\frac{(z+2)^2}{2z (z^2-16)} \frac{c_3}{r^{z+4}}, \quad
s_2^{(2,0)} = - \frac{(z+2)}{2z (z-6)} c_3 r^{z-6}.
\end{equation}
For the functions $F^{(1,1)}$, there is in principle a source term in
(\ref{fj1}-\ref{kl}), but it involves the combination
\begin{equation}
r k_T^{(0,1)'} + rs_1^{(1,0)'} - r^{-2(z-1)} rs_2^{(1,0)'},
\end{equation}
which vanishes by virtue of the equation of motion for $k_T^{(0,1)}$,
\eqref{kt01}. Thus, the particular integral will only involve
$k_T^{(1,1)}$, $s_1^{(1,1)}$ and $s_2^{(1,1)}$, with powers of $r$
such that the resulting particular integral makes no contribution to
the stress tensor.

Considering the stress tensor for the solutions proportional to $c_3$,
we see that there are potentially divergent contributions to
${\mathcal E}_x$ coming from $s_1^{(l,0)}$ and $s_2^{(l,0)}$ for $l <
z$. However, for this mode (recall again that there is no source for
$k_L$ at this order and so no $k_L$ in the formula below)
\begin{equation}
{\mathcal E}_x = r^{z+2} \omega k \left[ r s_2' + \frac{(z-2)}{z}
  r^{2z-1} s_1' + \frac{k^2}{2zr^2} (r s_2' - r^{2z-1} s_1' - 2(z-1)
  s_2) \right]e^{i \omega t + i kx},
\end{equation}
and this will vanish by virtue of the $\omega^0$ part of
\eqref{vec1}. This is not surprising; having learnt that there are no
divergent contributions to $\mathcal E$, a divergent contribution to
${\mathcal E}_x$ would be incompatible with the energy conservation
equation. We can see this explicitly at the first two orders in $k^2$
using the $s_i^{(1,0)}$ and $s_i^{(2,0)}$ calculated
above.

The contribution to the stress tensor from the solution parametrized
by $c_3$ is then
\begin{equation}
{\mathcal P}_x = 2 \frac{(z-1)(z+2)}{z} \omega k^3 c_3 e^{i \omega t + i
  kx}, \quad \Pi_{xx}
= - \Pi_{yy} = - 2 \omega^2 k^2 \frac{(z-1)(z+2)}{z}  c_3 e^{i \omega t + i kx}.
\end{equation}
Note that the conservation equation $\omega {\mathcal P}_x + k
\Pi_{xx} = 0$ is satisfied.

In summary, in the scalar sector, we have a three-parameter family of
solutions of the equations of motion which satisfy our asymptotic
boundary conditions. The stress tensor only depends on two of the
parameters, and is finite and conserved, with all the components we
would expect;
\begin{equation}
{\mathcal E} = k c_2'e^{i \omega t + i kx}, \quad {\mathcal E}_x = -
\omega c_2'e^{i \omega t + i kx},
\end{equation}
\begin{equation}
{\mathcal P}_x = k c_3'e^{i \omega t + i kx}, \quad \Pi_{xx} = -\omega
c_3'e^{i \omega t + i kx}, \Pi_{yy} = (z k c_2'
+ \omega c_3') e^{i \omega t + i kx},
\end{equation}
where to simplify the form of the stress tensor we write $c_2' = 2
\frac{(z+2)}{z} k c_2$ and $c_3' = 2 \frac{(z-1)(z+2)}{z} \omega k^2 c_3$.

\subsubsection{Vector modes}

Consider now the vector modes, described by the functions
$v_1(r), v_2(r), v_3(r)$. The equations of motion for these are
\begin{eqnarray}
\omega (r v_1' - r^{-2(z-1)} rv_2') &=& -k^2 r v_3', \label{vcons} \\
r^2 v_1'' + (2z+1) r v_1' + z r^{-2(z-1)} r v_2' &=& \left(
  \frac{k^2}{r^2} - \frac{\omega^2}{r^{2z}} \right) v_1, \\
r^2 v_3'' + (z+3) r v_3' + \omega \frac{v_1}{r^2} - \omega
\frac{v_2}{r^{2z}} &=& -\frac{\omega^2}{r^{2z}} v_3.
\end{eqnarray}

For the vector part, the non-zero parts of the stress tensor complex are
\begin{equation}\label{st_vec1}
{\mathcal E}_y = r^{z+2} \left[ r \partial_r v_{2} +
  \frac{(z-2)}{z} r^{2(z-1)} r \partial_r v_{1}\right] e^{i\omega
  t+ i k x} + {\mathcal E}_y^{deriv} ,
\end{equation}
\begin{equation}\label{st_vec2}
{\mathcal P}_y = r^{z+2} [-r \partial_r v_{1} + r^{-2(z-1)}
r \partial_r v_{2} ] e^{i\omega t+ i k x} + {\mathcal P}_y^{deriv} ,
\end{equation}
\begin{equation}\label{st_vec3}
\Pi_{xy} = - r^{z+2} k  r \partial_r v_3  e^{i\omega
  t+ i k x} + \Pi_{xy}^{deriv}.
\end{equation}
The first equation \eqref{vcons} imposes the conservation equation
$\omega {\mathcal P}_y + k \Pi_{xy} = 0$.

Note that if $\omega=0$, the first equation implies that $v_3'=0$, and
$v_3$ drops out of the system of equations---it vanishes up to a
possible constant term. We will drop constant terms in $v_1, v_2$ and
$v_3$ as not satisfying the asymptotic boundary conditions. Therefore
$v_3$ will vanish if $\omega =0$, so we rescale $v_3 \to \omega
v_3$. Then the equations of motion are
\begin{eqnarray}
r v_1' - r^{-2(z-1)} rv_2' &=& -k^2 r v_3', \\
r^2 v_1'' + (2z+1) r v_1' + z r^{-2(z-1)} r v_2' &=& \left(
  \frac{k^2}{r^2} - \frac{\omega^2}{r^{2z}} \right) v_1, \\
r^2 v_3'' + (z+3) r v_3' + \frac{v_1}{r^2} -
\frac{v_2}{r^{2z}} &=& -\frac{\omega^2}{r^{2z}} v_3.
\end{eqnarray}
We again solve these equations perturbatively
in $k^2$, $\omega^2$, writing $F = \sum_{l,m} k^{2l} \omega^{2m}
F^{(l,m)}$ and treating the RHS as a source term for the
solution at a given order determined in terms of the solution at
earlier orders. At the leading order, the solution is
\begin{equation}
v_1^{(0,0)}(r) = \frac{c_4}{r^{3z}}, \quad v_2^{(0,0)}(r) = \frac{3z}{z+2}
\frac{c_4}{r^{z+2}}, \quad v_3^{(0,0)}(r) = \frac{c_5}{r^{z+2}} +
\frac{(z-1)}{z(z+2)(3z+2)}  \frac{c_4}{r^{3z+2}},
\end{equation}
where we have once again set constant terms to zero by the boundary
conditions. In terms of the stress tensor, the constant $c_4$ is
associated with a finite contribution to ${\mathcal E}_y$, and $c_5$
is associated with a finite contribution to $\Pi_{xy}$. To evaluate
the full stress tensor, we need some of the higher-order terms.

In this case, the equations are the same at each order, so there are
no new homogeneous solutions; homogeneous solutions at higher order
can be absorbed into a redefinition of $c_4, c_5$. We therefore need
to consider only relevant particular integrals.  The particular
integrals from the solution parametrized by $c_4$ make no contribution
to the stress tensor. However, the particular integrals $v_1^{(l,0)}$,
$v_2^{(l,0)}$ for $l < z$ associated to the solution parametrized by
$c_5$ will make potentially divergent contributions to ${\mathcal
  E}_y$. As in the constant perturbations, for $z \geq 4$, we will
need to set $c_5=0$ to satisfy the boundary condition $v_2 \to 0$ as
$z \to \infty$.  The divergences then involve the particular integrals
up to $l=3$, which are
\begin{equation}
v_1^{(1,0)} = - \frac{c_5 z}{2(z-1)r^{z+2}}, \quad v_2^{(1,0)} = -
\frac{c_5 (z^2-4)r^{z-4}}{2(z-4)(z-1)}, \quad v_3^{(1,0)} = -
\frac{c_5}{(z^2-16)r^{z+4}},
\end{equation}
\begin{equation}
v_1^{(2,0)} = \frac{3 c_5 z}{4(z-1)(z^2-16)r^{z+4}}, \quad v_2^{(2,0)} =
\frac{c_5 r^{z-6}}{4(z-6)(z-1)}, 
\end{equation}
\begin{equation}
v_3^{(2,0)} = - \frac{c_5(z-8)}{8(z^2-16)(z^2-36)r^{z+6}},
\end{equation}
\begin{equation}
v_1^{(3,0)} = \frac{c_5 z(z-11)}{16(z-1)(z-3)(z+4)(z^2-36)r^{z+6}},
\quad v_2^{(3,0)} =  \frac{c_5(z^2-3z+8)r^{z-8}}{2(z-3)(z-8)(z^2-16)},
\end{equation}
\begin{equation}
v_3^{(3,0)} = \frac{c_5(5z^2-43z+72)}{24(z-3)(z^2-16)(z^2-36)(z^2-64)r^{z+8}}.
\end{equation}
Note that at $z=3$, this form for the particular integral will not
apply, and it will be replaced by a solution involving logarithms, as
occurred for $z=2$ in the scalar sector. We have not determined this
solution explicitly as this is not a particularly interesting value of
$z$. We also need to consider the particular integral $v_i^{(0,1)}$
for the solution parametrized by $c_5$, as the contribution to $v_2$
would go like $r^{-(z+2)}$, and hence could make a finite
contribution. A particular integral is
\begin{equation}
v_1^{(0,1)} = 0, \quad v_2^{(0,1)} = 0, \quad
v_3^{(0,1)} = -\frac{c_5}{2z(3z+2)r^{3z+2}},
\end{equation}
so this will make no contribution to the stress tensor.

We can now use this to
calculate the value of the contribution to the stress tensor from
this mode for generic $z$. We have
\begin{eqnarray} {\mathcal E}_y &=&  r^{z+2} \left[ r \partial_r v_{2}
    + \frac{(z-2)}{z} r^{2(z-1)} r \partial_r v_{1}
  \right]e^{i\omega t +ikx} + {\mathcal E}_y^{deriv}
  \\ &=&  \nonumber-[ 6(z-1) c_4 + \frac{c_5
  }{2(z-4)} r^{2z-4} k^4 - \frac{(z-5) c_5}{2(z-3)(z-6)(z^2-16)}
r^{2z-6} k^6] e^{i\omega t+ikx}+ {\mathcal E}_y^{deriv},
\end{eqnarray}
while from appendix \ref{deriv}, we have
\begin{eqnarray}
{\mathcal E}_y^{deriv} &=& c_5 \left[ \left( -\frac{2 \sigma_1}{(z-4)} -
    \sigma_2 \right) k^4 r^{2z-4} \right. \\ && \left. + \left( - \frac{(z-8) \sigma_1}{2
        (z-6)(z^2-16)} + \frac{3 \sigma_2}{2(z^2-16)} - \sigma_3 \right)
  k^6 r^{2z-6} \right]e^{i\omega t +ikx}. \nonumber
\end{eqnarray}
The term at order $k^4$ grows at large $r$ for $z >2$, and the
term at order $k^6$ grows at large $r$ for $z >3$. Since we can only
consider this mode for $z < 4$, there are no further divergences.
We can cancel the divergent terms by setting
\begin{equation}
\sigma_2 = -\frac{4\sigma_1+1}{2(z-4)}, \quad \sigma_3 =
-\frac{(z-3)(7z-44) \sigma_1 +(z-1)(z-8)}{4(z-3)(z-4)(z-6)(z^2-16)}.
\end{equation}
Thus, for generic $z$, we can obtain a finite stress
tensor complex in the linearised approximation by choosing appropriate
curvature counterterms in our definition of the action.\footnote{For $z=3$,
a different choice of coefficients will be required.} With this
choice of action, 
\begin{equation} 
{\mathcal E}_y = - 6(z-1) c_4  e^{i\omega t+ikx}.
\end{equation}

For the other components, we have
\begin{equation}
{\mathcal P}_y = r^{z+2} [-r \partial_r v_{1} + r^{-2(z-1)}
r \partial_r v_{2} ] e^{i\omega t +ikx} + {\mathcal P}_y^{deriv}  =
-(z+2) c_5 k^2e^{i\omega t +ikx},
\end{equation}
and
\begin{equation}
  \Pi_{xy} = - k \omega r^{z+2}
  r \partial_r v_3 e^{i\omega t+ikx} + \Pi_{xy}^{deriv} =  (z+2) c_5 k
  \omega e^{i\omega t+ikx} .
\end{equation}
We see in appendix \ref{deriv} that the curvature components make no
contributions to these components.  As a consistency check, we see
that $\omega {\mathcal P}_y + k \Pi_{xy} = 0$.

In summary, for the non-zero momentum perturbations, we have a
five-parameter family of solutions (parametrized by $c_1$, $c_2$ and
$c_3$ in the scalar sector and $c_4$ and $c_5$ in the vector
sector). For $z < 2$, in the scalar sector, we imposed a tighter
boundary condition by setting $d_3=0$. For the non-zero momentum
perturbations, this is required to get a finite energy density, and
hence to satisfy $\delta S=0$. Four of the parameters correspond to
the independent components of the stress tensor in this non-zero
momentum sector; as in the previous constant case, the scalar
mode parametrized by $c_1$ does not contribute to the stress tensor
complex at this linear order. For $z \geq 4$, we must again set
${\mathcal P}_i = 0$, and we are left with a three-parameter family of
solutions.

Note that as we said earlier, for $z \geq 2$, this linearised
calculation of the relation between the asymptotic behaviour of the
metric and the action will not be reliable for a general
perturbation. However, this calculation is always applicable if we
consider a specific case where the bulk spacetime is everywhere a
small perturbation away from the background. It is thus a significant
result that can obtain a completely finite stress tensor complex at
this linearised level by an appropriate choice of counterterms in the
action.

\subsection{Vanishing of the variation of the action}
\label{vanish}

We return briefly to the question of the vanishing of the variation of
the action. We have seen that all components of the stress tensor are
finite. Thus, $\delta S=0$ if $\delta \hat{e}^A_\alpha \to 0$ as $r
\to \infty$.  The coefficient of the variation of the vector field is
\begin{equation}
  s_0 = - \alpha [z r^{z+2} \hat{a}_t + r^{z+2} r \partial_r
  (\frac{1}{2} \hat{h}_{tt} + \hat{a}_t) - r^2 \partial_t \hat{a}_r ]
  + s_0^{deriv}.
\end{equation}
Since $\hat{h}_{tt}$ and $\hat{a}_t$ have components that go like
$r^{-\frac{1}{2}(z+2+\beta_z)}$, $s_0$ will have a divergence like
$r^{\frac{1}{2}(z+2-\beta_z)}$, which gives a positive power of $r$
for $1 \leq z < 2$. (For $z=2$, this is replaced by a $\ln r$
divergence). However, it is precisely for this range that we impose the
stronger boundary condition, which implies that $\delta A^0$ vanishes
more quickly than $r^{-\frac{1}{2}(z+2-\beta_z)}$ as $r \to
\infty$. This is precisely what is required to ensure that the $s_0
\delta A^0$ contribution also vanishes, so we indeed satisfy $\delta S
=0$ on-shell.

Thus, for our asymptotic boundary conditions, \eqref{action} is a good
action principle for the asymptotically Lifshitz spacetimes, as it is
finite on-shell and satisfies $\delta S=0$ for arbitrary variations
satisfying the boundary conditions.
\subsection{Operator dual to $A^0$}

We have shown that the stress tensor is finite for our action. We
should also consider the operator dual to $A^0$, and see if its
expectation value is finite. As we remarked earlier, for $1 \leq z <
2$, it seems natural to think of the part of $A^0$ falling off as
$r^{-\frac{1}{2}(z+2-\beta_z)}$ as a non-normalizable mode (that is,
as boundary data associated with the vector field). If we write
$\delta A^0 = r^{-\frac{1}{2}(z+2-\beta_z)} \delta \bar{A}^0$,
$\bar{s}_0 = r^{\frac{1}{2}(z+2-\beta_z)} s_0$ is the coefficient of
$\delta \bar{A}^0$ in the variation of the action, which would be
interpreted as the expectation value of the dual operator. The term
falling off as $r^{-\frac{1}{2}(z+2+\beta_z)}$ makes a finite
contribution to $\bar{s}_0$, so it can be thought of as the
corresponding normalizable mode. In our linearised analysis, this
implies that the additional scalar mode which does not contribute to
the stress tensor can be interpreted as the expectation value of the
operator dual to changes in the non-normalisable mode for the vector
field.

However, it is difficult to extend this analysis to $z \geq 2$. The
mode which falls off like $r^{-\frac{1}{2}(z+2-\beta_z)}$ then
violates our boundary conditions for the metric components, so it is
not clear if it can still be interpreted as boundary data for the
vector field. If we calculate $\bar{s}_0$ anyway, it has a finite
contribution from the mode which falls off as
$r^{-\frac{1}{2}(z+2+\beta_z)}$, which suggests this mode can be given
the same interpretation, but it now also has a divergent contribution
from the mode which falls off as $r^{-(z+2)}$. If we want to think of
the mode in our linearised analysis which falls off as
$r^{-\frac{1}{2}(z+2-\beta_z)}$ as the boundary data we are varying,
then because this mode appears in $f$ and $k$ as well as $j$, the
coefficient of this variation is really a linear combination of
$\bar{s}_0$, ${\mathcal E}$ and $\Pi_i^i$ (this didn't make any
difference for $1 \leq z < 2$ because the contribution from the mode
which falls off as $r^{-(z+2)}$ vanished). However, this does not seem
to cancel the divergence. There is a combination of
$\bar{s}_0$, ${\mathcal E}$ and $\Pi_i^i$ which will cancel the
divergent contribution from the mode which falls off as $r^{-(z+2)}$,
but the coefficients are different from those implied by the solution
of the linearised equations. We leave the resolution of
this conundrum for future work.

\section{Discussion}
\label{disc}

The main results of this paper are that first, we have constructed an
appropriate action principle for asymptotically Lifshitz spacetimes
with a flat boundary in the massive vector theory of
\cite{taylor}. We then proposed a definition of the non-relativistic
stress tensor complex for both the Schr\"odinger and Lifshitz cases in
terms of the variation of the action. Our proposal corresponds to the
proposal of \cite{him} in the relativistic case, taking the
appropriate variation to be a variation of the boundary frame fields
holding the matter fields with tangent space indices fixed. This is
one of our key results: the major difference between the calculation
of the stress tensor in these cases and the more familiar AdS case is
not the different scaling of different directions, but simply the fact
that we need to take the contribution to the stress tensor from
variation of the vector field into account. Once we have correctly
accounted for this, we get finite answers for the stress tensor
complex.

In the Schr\"odinger case, we have shown that this proposal agrees
with the stress tensor complex obtained by re-interpreting the stress
tensor of the related asymptotically AdS spacetime in terms of the
non-relativistic field theory \cite{mmt}, for asymptotically
Schr\"odinger spacetimes which can be obtained by TsT transformation
from a vacuum asymptotically AdS spacetime. We expect this will be
true in general, but have left the detailed calculation for future
work. In the Lifshitz case, we have solved the linearised equations of
motion for the general perturbation about the background
\eqref{lmet}. This enables us to relate the stress tensor to the
asymptotic falloff of the metric and vector fields of an
asymptotically Lifshitz spacetime. We have shown that the resulting
stress tensor is finite.

There are a number of interesting directions for future work.  For the
Schr\"odinger case, it would be useful to establish the minimal
boundary conditions for which we have a well-defined action principle,
parallelling our analysis for the Lifshitz case. Our results on
finiteness of the stress tensor imply that we can relax the boundary
conditions somewhat relative to those used in \cite{heat}, but as in
the Lifshitz case, there are divergences in the matter sector that
need to be addressed. There has been extensive work on obtaining
Schr\"odinger geometries in different contexts
\cite{Kovtun:2008qy,Hartnoll:2008rs,Colgain:2009wm,Bobev:2009mw,Donos:2009xc},
and it would be useful to work out the boundary counterterms required
to construct appropriate action principles in these different
cases. The fact that we now have a proposal for constructing the
stress tensor directly in the asymptotically Schr\"odinger solution is
particularly useful for these cases where we do not have a
solution-generating transformation relating asymptotically AdS and
asymptotically Schr\"odinger solutions.

For the asymptotically Lifshitz geometries, to complete the analysis
of one-point functions, we need to resolve the problems with the
calculation of the expectation value for the operator dual to the
vector field raised in the previous subsection. It would also be
interesting to use our action to calculate two-point functions in the
pure Lifshitz background. In the study of finite-temperature
geometries, it would be interesting to further pursue the holographic
renormalization framework by understanding the construction of more
general black hole solutions corresponding to arbitrary hydrodynamic
stress tensors from the dual field theory point of view. A natural
next step is to construct a black hole with non-zero spatial
momentum. Because the background geometry does not have a boost
invariance, this cannot be obtained by simply boosting the known
solutions.

Lifshitz points usually occur at the juncture of three phase
boundaries. There is more than one ordered phase below a critical
temperature. Depending on an external control parameter, for instance
an external field, one can make a transition from an ordered phase
where the condensate is spatially uniform to a new ordered phase where
the order is inhomogeneous; that is, the critical system is allowed to
have a phase where the Landau energy of the system is minimized by a
non-uniform condensate rather than a homogeneous one \cite{Luban}.
Lifshitz critical points are relevant for studying interesting
condensed matter systems including superconductors and Liquid Crystals
among others \cite{Schuh} and \cite{Liquid}. Recently hairy black
holes in AdS have been of central importance in modeling second order
transitions in the context of AdS/CFT (notably
superfluid/superconductor transition). The second order transition was
modeled by a charged scalar condensing in the vicinity of a black hole
event horizon \cite{Gubser2008, Hartnoll_Horowitz} in AdS. Perhaps the
first step to model a Lifshitz point at finite temperature is to study
a similar set up but with the bulk black hole replaced with a Lifshitz
black hole; one should see if a hairy Lifshitz black hole with
$z\neq1$ can be constructed.

For both cases, it would be interesting to extend the analysis to
consider more general boundary data. We have restricted ourselves to
the case where the boundary is flat, but a similar definition of the
stress tensor can be applied for a general curved spatial metric
$g_{ij}$. The most interesting case to consider is when the boundary
metric is a sphere. This introduces additional slow falloff terms in
the asymptotics, so we would need to check again that the resulting
stress tensor is finite, and hence that we have a well-defined action
principle for such boundary conditions. The perturbation analysis for
asymptotically Lifshitz spacetimes with a spherical boundary has been
initiated in \cite{dt,peet}.

At a more formal level, we would like to have a better understanding
of the possible boundary data for asymptotically Lifshitz spacetimes.
In particular, there are issues we have not yet fully understood about
the meaning of our calculation of the energy flux from the point of
view of a non-relativistic theory. By analogy with the relativistic
case, we have constructed our stress tensor by considering arbitrary
variations of the boundary data, including variations $\delta
\hat{e}^{(0)}_i$, which give the energy flux. However, introducing such
components does not seem natural from the point of view of a
non-relativistic theory. In a non-relativistic theory, the flat
background spacetime we considered above can be thought of as a fiber
bundle, with the spatial slices fibered over the time direction
\cite{penrose}. Each spatial slice corresponds to a moment in time,
and relative position in the spatial slices is invariantly defined,
but there is no invariant notion of relative position in different
spatial slices. Allowing components $\hat{e}^{(i)}_t$ is consistent with
this fiber bundle structure, but it is not clear how $\hat{e}^{(0)}_i$
would be.  It would be interesting to understand this distinction
between the different components of the stress tensor complex more
fully.

Another general issue is to find a truncation of string theory which
gives a Lifshitz geometry with anisotropic scaling symmetry. That is,
where the metric takes the form \eqref{lifmet}, and the matter fields
are also invariant under the isometry $t \to \lambda^z t$, $x^i \to
\lambda x^i$, $r \to \lambda^{-1} r$. The symmetry implies that any
scalar field must be a constant, which makes it difficult to find an
embedding in string theory, where a timelike vector field is usually
accompanied by a non-trivial scalar.

An important general issue for applications of holography to condensed
matter systems is that it is not generally understood what the
conditions are under which the theory has a classical weakly curved
gravitational dual. That is, what is the analogue of the large $N$
limit for gauge theories which implies that quantum corrections to the
gravity theory under control?

\section*{Acknowledgements}

We are grateful for useful discussions with Simon Caron-Huot, Veronika
Hubeny, Alex Maloney, Don Marolf and Mukund Rangamani, and thank Don
Marolf and Mukund Rangamani for reading a draft of this paper. SFR is
supported by STFC. OS is supported by NSERC of Canada.

\appendix

\section{Derivative contributions to the Lifshitz stress tensor}
\label{deriv}

In this appendix, we will evaluate the contributions of the part of
the boundary action involving derivatives to the stress tensor for the
asymptotically Lifshitz spacetimes in our linearised perturbative
analysis. We will not discuss the most general possible derivative
terms, but consider a simple set of terms up to fourth order in
derivatives which are sufficient to cancel the divergences in
${\mathcal E}_y$, giving us a finite stress tensor complex for the
linearised perturbations. The form of the action will not be uniquely
fixed by imposing finiteness of the stress tensor; our aim here is
simply to show that there is a choice for the counterterms $S_{deriv}$
which gives a finite answer for the stress tensor.
We consider an action
\begin{equation}
  S_{deriv} = \frac{1}{16 \pi G_4} \int d^3 \xi \sqrt{-h}  [\sigma_1 R^h
  + \sigma_2 \nabla_\alpha A_\beta \nabla^\alpha A^\beta + \sigma_3 (\Box
  A_\alpha)(\Box A^\alpha)],
\end{equation}
where $R^h$ is the curvature of the boundary metric, and the
$\sigma_i$ are arbitrary constants.

We get no contribution from $S_{deriv}$ for constant
perturbations. For the non-zero momentum modes considered in section
\ref{gpert}, the contribution to the boundary stress tensor complex
becomes
\begin{eqnarray} \label{glcurv}
{\mathcal E}^{deriv} &=& -r^{z} k^2 [\sigma_1 (k_L - k^2 k_T) + 2
\alpha^2 \omega (\sigma_2 - \sigma_3 \Box) s_1] e^{i\omega
  t+ i k x}, \\ \nonumber
{\mathcal E}_x^{deriv} &=&  r^{z} k \omega [\sigma_1  (k_L - k^2
k_T)  + 2 \alpha^2 \omega (\sigma_2 - \sigma_3 \Box) s_1] e^{i\omega
  t+ i k x}, \\ \nonumber
{\mathcal E}_y^{deriv}  &=& -\left[ \sigma_1 (r^z (k^2 \omega v_3 - k^2
v_2) + r^{3z-2} k^2 v_1) - \alpha^2 r^{3z} \left( \frac{k^2}{r^2} - 2
  \frac{\omega^2}{r^{2z}} \right) (\sigma_2 - \sigma_3 \Box)
v_1 \right]e^{i\omega t+ i k x}, \\ \nonumber
{\mathcal P}_x^{deriv} &=&  r^{z} k [\sigma_1 r^{-2z+2} \omega (k_L - k^2
k_T) + 2\alpha^2 k^2 (\sigma_2 - \sigma_3 \Box) s_1] e^{i\omega t+ i k
  x}, \\ \nonumber
{\mathcal P}_y^{deriv} &=& -r^z k^2 [\sigma_1 ( r^{-2z+2} (\omega v_3 -
v_2) + v_1 ) + \alpha^2 (\sigma_2 - \sigma_3 \Box) v_1] e^{i\omega t+
  i k x}, \\ \nonumber
\Pi_{xx}^{deriv} &=& -r^{z} \omega [\sigma_1 r^{-2z+2} \omega (k_L - k^2
k_T) + 2\alpha^2 k^2 (\sigma_2 - \sigma_3 \Box) s_1] e^{i\omega t+ i k
  x},  \\ \nonumber
\Pi_{xy}^{deriv} &=& r^z k \omega [\sigma_1 ( r^{-2z+2} (\omega v_3 -
v_2) + v_1 ) + \alpha^2 (\sigma_2 - \sigma_3 \Box) v_1]e^{i\omega t+ i
  k x}, \\ \nonumber
\Pi_{yy}^{deriv} &=& \sigma_1 [ r^z (-k^2 f -2 k^2 \omega s_1) +
r^{2-z} (2 k^2 \omega
s_2 - \omega^2 k_L - k^2 \omega^2 k_T) ] e^{i\omega t+ i k x},
\end{eqnarray}
where for the plane wave perturbations, $\Box =
\frac{\omega^2}{r^{2z}} - \frac{k^2}{r^{2}}$.  We can see immediately
that this contribution to the stress tensor is separately conserved,
as we would expect: $\omega {\mathcal E}^{deriv} + k {\mathcal
  E}_x^{deriv} = 0$, $\omega {\mathcal P}_x^{deriv} + k \Pi_{xx}^{deriv} = 0$,
$\omega {\mathcal P}_y^{deriv} + k \Pi_{xy}^{deriv} = 0$. Note that this did
not require the use of the equations of motion, unlike for the part of
the  action we treated in the body of the paper.

The contributions to most components of the stress tensor complex from
the derivative terms will vanish. The general point is that the
derivative terms are suppressed relative to the terms considered
earlier by factors of $k^2/r^2$ or $\omega^2/r^{2z}$. Hence when the
earlier terms give finite contributions, the derivative terms will
give vanishing contributions. Explicitly, the scalar components
${\mathcal E}^{deriv}, {\mathcal E}_x^{deriv}, {\mathcal P}_x^{deriv},
\Pi_{xx}^{deriv}$, and $\Pi_{yy}^{deriv}$ involve $r^z f$, $r^z k_L$,
$r^z k_T$, $r^z s_1$ and $r^z s_2$, (or smaller powers of $r$) all of
which vanish for the general solution of the linearised equations
satisfying our boundary conditions obtained in section
\ref{gpert}. Similarly, for the vector sector, the components
${\mathcal P}_y^{deriv}$ and $\Pi_{xy}^{deriv}$ involve $r^{-z+2}
v_3$, $r^{-z+2} v_2$ and $r^z v_1$, all of which vanish for the
general solution of the linearised equations satisfying our boundary
conditions obtained in section \ref{gpert}.

The one exception is ${\mathcal E}_y$, which involves $r^z v_3$, $r^z
v_2$, and $r^{3z-2} v_1$, which vanish for the solution parametrized
by $c_4$, but not for that parametrized by $c_5$. This is precisely
where we found divergences for the terms coming from the
non-derivative part of the action, so we want to evaluate the
derivative terms and see that we can choose the coefficients to cancel
these divergences. It is the $v_1$ and $v_2$ terms which produce
potential divergences; there is a term which goes like $k^4 r^{2z-4}$
from putting $v_i^{(1,0)}$ in the $\sigma_1, \sigma_2$ terms, and
terms that go like $k^6 r^{2z-6}$ from putting $v_i^{(2,0)}$ in the
$\sigma_1, \sigma_2$ terms and from putting $v_i^{(1,0)}$ in the
$\sigma_3$ term. As we only have a $c_5$ mode for $z < 4$, these are
the only potential finite or divergent terms. Putting them together,
the divergent terms for this mode are
\begin{eqnarray}
{\mathcal E}_y^{deriv} &=& c_5 \left[ \left( -\frac{2 \sigma_1}{(z-4)} -
    \sigma_2 \right) k^4 r^{2z-4} \right. \\ && \left. + \left( - \frac{(z-8) \sigma_1}{2
        (z-6)(z^2-16)} + \frac{3 \sigma_2}{2(z^2-16)} - \sigma_3 \right)
  k^6 r^{2z-6} \right]e^{i\omega t +ikx}. \nonumber
\end{eqnarray}
The terms we are omitting in this expression vanish as $r \to \infty$
for $z <4$. We can then choose the coefficients $\sigma_i$ to cancel
these divergences against the divergences in ${\mathcal E}_y$ from the
non-derivative part of the action; we do this explicitly in section
\ref{gpert}. Since there are two divergences to cancel and three
coefficients, this will not fix the form of the action uniquely. Thus,
there is an action for which the stress tensor is finite, but
this condition does not determine a unique choice of the action. We
kept two terms at second order in derivatives in our discussion of
$S_{deriv}$ to illustrate this failure to fix a unique action.

If we considered more general boundary data, such as where the spatial
metric is replaced by a sphere, there will be further constraints on
the coefficients in the derivative terms. For example, if we consider
a Lifshitz spacetime where the spatial sections are spheres, the term
involving the curvature of the boundary metric will contribute to the
action, but the terms involving derivatives of the vector field will
not. The coefficient of the curvature term can then be fixed by
cancelling the divergence in the on-shell action arising from the new
terms in the metric at relative order $1/r^2$. We leave a detailed
discussion of the extension of our analysis to more general boundary
data for future work.

\section{Euclidean Action and Thermodynamic Energy}
\label{Euclidean}

In this section, we show that our definition of the energy density for
asymptotically Lifshitz spacetimes agrees with the thermodynamic
energy density obtained by using the Euclidean version of the black
hole solution as a saddle-point in the path integral for a class of
static asymptotically Lifshitz black hole spacetimes. Our analysis in
this section will not use the linearised analysis we used previously;
we find that we can rewrite the action for the black hole solutions we
consider in an appropriate form just by using the equations of motion
(\ref{Eeqn},\ref{Meqn}).

We consider a metric ansatz
\begin{equation}\label{ansatz}
ds^2=-p(r)dt^2+q(r)(dx^2+dy^2)+\frac{dr^2}{r^2}, \quad A_t=A_t(r).
\end{equation}
This will give a black hole solution if there is an event horizon at
$r=r_H$, where $p(r) = p_H(r-r_H)^2 + {\mathcal O}(r-r_H)^3$. For a
regular horizon, we must also have $A_t(r) = A_{tH} (r-r_H) + {\mathcal
  O}(r-r_H)^3$. We assume that there is a solution of the equations of
motion with these properties; such solutions were constructed
numerically in \cite{dt,mann}.

If we rotate $t \to -i \tau$, the Euclidean black hole solution gives a
saddle-point approximation to the path integral defining the thermal
partition function at temperature
\begin{equation}\label{temp}
T_{H}=r_{H}\frac{\sqrt{p_H}}{2\pi}.
\end{equation}
In the Euclidean black hole solution, the radial coordinate is
restricted to $r_H \leq r < \infty$, with a smooth origin at $r=r_H$
once we choose $\Delta \tau = \beta = T_H^{-1}$. The action for this
black hole solution gives an approximation to the free energy, $F =
T_H I_{Eucl}$.  Since the solution is translationally invariant in $x$
and $y$, it is natural to divide by the coordinate volume in those
directions to define the free energy density $f = T_H
I_{Eucl}/V_2$. The entropy density is given by the area of the black
hole horizon,
\begin{equation}\label{entropy}
s=\frac{A}{4G_4}=\frac{q(r_{H})}{4G_4},
\end{equation}
so we can define the thermodynamic energy density by
\begin{equation}
f = {\mathcal E}_{thermo} - T_H s.
\end{equation}
We want to see that this agrees with the energy density we defined previously,
${\mathcal E}_{thermo} = {\mathcal E}$.

To do so, we use the equations of motion to rewrite the on-shell
Euclidean action (\ref{action}) in terms of boundary terms at the
asymptotic boundary and at the horizon.

After the analytic continuation $t \to -i\tau$, the action of the
Euclidean solution is
\begin{eqnarray} \label{eaction}
I^E &=& -\frac{1}{16 \pi G_4} \int d^4x \sqrt{g} (R - 2\Lambda -
\frac{1}{4} F_{\mu\nu} F^{\mu\nu} - \frac{1}{2} m^2 A_\mu A^\mu) \\ &&-
\frac{1}{16 \pi G_4} \int d^3 \xi \sqrt{h} (2K - 4 - z\alpha \sqrt{ -
  A_\alpha A^\alpha} )- I_{deriv}. \nonumber
\end{eqnarray}
To relate the action to the boundary terms, it is convenient to use
the equation of motion for the vector field \eqref{Meqn} to write $m^2
A_\mu A^\mu + \frac{1}{2} F_{\mu\nu} F^{\mu\nu} = \nabla_\mu
(F^\mu_\nu A^\nu)$, so
\begin{eqnarray}
I^E &=& -\frac{1}{16 \pi G_4} \int d^4x \sqrt{g} (R - 2\Lambda +
\frac{1}{4} F_{\mu\nu} F^{\mu\nu} + \frac{1}{2} m^2 A_\mu A^\mu) \\ &&+
\frac{1}{16 \pi G_4} \int d^3 \xi \sqrt{h} (n^\mu F_{\mu\nu} A^\nu -
2K + 4 + z\alpha \sqrt{ -
  A_\alpha A^\alpha} )- I_{deriv}. \nonumber
\end{eqnarray}
Now for the ansatz \eqref{ansatz}, the derivative terms do not
contribute and the only non-zero component of $F_{\mu\nu}$ is $F_{rt}
= -F_{tr}$, so $A_t A^t = A_\mu A^\mu$ and $F_{rt} F^{rt} =
\frac{1}{2} F_{\mu\nu} F^{\mu\nu}$, and hence the Einstein equations
\eqref{Eeqn} imply
\begin{equation}
R^x_{\ x} + R^y_{\ y} + R^r_{\ r} - R^t_{\ t} = 2 \Lambda -
\frac{1}{4} F_{\mu\nu} F^{\mu\nu} - \frac{1}{2} m^2 A_\mu A^\mu.
\end{equation}
Thus, the on-shell action for the ansatz \eqref{ansatz} is
\begin{equation}
I^E = -\frac{1}{16 \pi G_4} \int d^4x \sqrt{g} 2R^t_t +
\frac{1}{16 \pi G_4} \int d^3 \xi \sqrt{h} (n^\mu F_{\mu\nu} A^\nu -
2K + 4 + z\alpha \sqrt{ -
  A_\alpha A^\alpha} ).
\end{equation}
Furthermore, using the form of the metric \eqref{ansatz}, we can show
\begin{eqnarray}
\sqrt{g}R^{t}_{~t}=-(\sqrt{h}K^{t}_{~t})^{'},
\end{eqnarray}
so the integration over $r$ can be rewritten in terms of boundary
terms. The integration over $\tau$ and $x,y$ gives an overall factor
of $\beta V_2$, which we divide out. Thus,
\begin{eqnarray}\label{I_tot}
16\pi G_4 \frac{I^{E}}{\beta
V_2}&=&\sqrt{h}(2K^{t}_{~t}+n^{\mu}F_{\mu\nu}A^{\nu}-2K+4+z\alpha\sqrt{-A_{\mu}A^{\mu}})\mid_{r_{b}}\\&&-
2\sqrt{h}K^{t}_{~t}\mid_{r_{H}},
\end{eqnarray}
where $r_b$ and $r_{H}$ are the location of the boundary and the
horizon respectively. At the horizon,
\begin{equation}
2\sqrt{h}K^{t}_{~t}\vert_{r_{H}}=r\frac{q(r)p(r)^{'}}{\sqrt{p(r)}}\vert_{r_{H}}=2r_{H}\sqrt{p_H}q(r_{H})=16\pi
G_4 T_H s,
\end{equation}
so the surface term at the horizon reproduces the term $-T_H s$ in
the free energy density. The surface term at infinity is hence giving
the thermodynamic energy density. Now using (\ref{sab}), (\ref{sa}),
\begin{eqnarray}\label{energy}
\mathcal{E}=2s^{t}_{~t}-s^{t}A_{t}=\sqrt{-h}(2K^{t}_{~t}-2K+4+n^{\mu}F_{\mu}^{~\nu}A_{\nu}+z\alpha\sqrt{-A_{\mu}A^{\mu}})\vert_{r_{b}},
\end{eqnarray}
so the surface term at infinity in \eqref{I_tot} is precisely our
energy density; that is, ${\mathcal E}_{thermo} = {\mathcal E}$ as desired.

\bibliographystyle{utphys}
\bibliography{lifshitz}

\end{document}